\documentclass[11pt]{article}\usepackage[]{graphicx}\usepackage[]{xcolor}
\makeatletter
\def\maxwidth{ %
  \ifdim\Gin@nat@width>\linewidth
    \linewidth
  \else
    \Gin@nat@width
  \fi
}
\makeatother

\definecolor{fgcolor}{rgb}{0.345, 0.345, 0.345}

\usepackage{framed}
\makeatletter
\newenvironment{kframe}{%
 \def\at@end@of@kframe{}%
 \ifinner\ifhmode%
  \def\at@end@of@kframe{\end{minipage}}%
  \begin{minipage}{\columnwidth}%
 \fi\fi%
 \def\FrameCommand##1{\hskip\@totalleftmargin \hskip-\fboxsep
 \colorbox{shadecolor}{##1}\hskip-\fboxsep
     \hskip-\linewidth \hskip-\@totalleftmargin \hskip\columnwidth}%
 \MakeFramed {\advance\hsize-\width
   \@totalleftmargin\z@ \linewidth\hsize
   \@setminipage}}%
 {\par\unskip\endMakeFramed%
 \at@end@of@kframe}
\makeatother

\definecolor{shadecolor}{rgb}{.97, .97, .97}
\definecolor{messagecolor}{rgb}{0, 0, 0}
\definecolor{warningcolor}{rgb}{1, 0, 1}
\definecolor{errorcolor}{rgb}{1, 0, 0}
\newenvironment{knitrout}{}{} % an empty environment to be redefined in TeX

\usepackage{alltt}

\usepackage{fullpage}
\usepackage{amsmath,amssymb}
\usepackage{bbm} % for \mathbbm{1}
\graphicspath{{figure/}}

\usepackage{natbib}
\usepackage{framed}
\usepackage[labelsep=endash]{caption}
\newcounter{Figcount}
\newcounter{tempFigure}

\newcommand\proglang[1]{\textsf{#1}}
\newcommand\code[1]{\texttt{#1}}
\newcommand\pkg[1]{\texttt{#1}}
\newcommand\E{\mathsf{E}}

\newcommand\spatPomp{spatPomp}

\usepackage{booktabs}
\usepackage{makecell}
\usepackage{tikz,pgfplots}
\pgfplotsset{compat=1.16}
\usepgfplotslibrary{fillbetween}
\usetikzlibrary{patterns}
\usetikzlibrary{arrows,chains,matrix,positioning,scopes,fit,decorations.pathreplacing}
\tikzset{join/.code=\tikzset{after node path={%
\ifx\tikzchainprevious\pgfutil@empty\else(\tikzchainprevious)%
edge[every join]#1(\tikzchaincurrent)\fi}}}
\makeatother
\tikzset{>=stealth,every on chain/.append style={join},
         every join/.style={->}}
\tikzstyle{labeled}=[execute at begin node=$\scriptstyle,
   execute at end node=$]
\newcommand{\FixedLengthArrow}{1,0}

\usepackage{enumerate,alltt,xstring}
\usepackage[ruled,noline,linesnumbered]{algorithm2e}
\SetKwFor{For}{for}{do}{end}
\SetKwFor{While}{while}{do}{end}
\SetKwInput{KwIn}{input}
\SetKwInput{KwOut}{output}
\SetKwInput{KwCplx}{complexity}
\SetKwInput{KwIndices}{note}
\SetKwBlock{Begin}{procedure}{}
\DontPrintSemicolon

\usepackage{placeins}

\usepackage[sort&compress]{cleveref}

\crefname{figure}{Figure}{Figures}
\Crefname{figure}{Figure}{Figures}
\crefname{table}{Table}{Tables}
\Crefname{table}{Table}{Tables}
\crefname{equation}{Equation}{Equations}
\Crefname{equation}{Equation}{Equations}
\creflabelformat{equation}{#2#1#3}
\crefname{appendix}{Appendix}{Appendices}
\Crefname{appendix}{Appendix}{Appendices}
\crefname{algorithm}{Algorithm}{Algorithms}
\Crefname{algorithm}{Algorithm}{Algorithms}
\crefname{section}{Section}{Sections}
\Crefname{section}{Section}{Sections}
\crefname{AlgoLine}{line}{lines}
\Crefname{AlgoLine}{Line}{Lines}

\newcommand\vecTheta{\Theta}

\newcommand\Thetadim{D_\theta}
\newcommand\thetadim{d_\theta}
\newcommand\vectheta{\theta}

\newcommand\data[1]{#1^*}

\newcommand\RP{\hspace{0.15mm}\mathrm{RP}}
\newcommand\IVP{\hspace{0.15mm}\mathrm{IVP}}

\newcommand\ABF{ABF}

\newcommand\rep{i}
\newcommand\altRep{\tilde\rep}
\newcommand\Rep{\mathcal{I}}

\newcommand\unit{u}
\newcommand\altUnit{\tilde{u}}
\newcommand\Unit{U}
\usepackage[mathscr]{euscript}
\renewcommand\time{n}
\renewcommand\vec[1]{\boldsymbol{#1}}
\newcommand\altTime{\tilde{n}}
\newcommand\Time{N}
\newcommand\Np{J}
\newcommand\np{j}
\newcommand\altNp{q}

\newcommand\resampleIndex{r}
\newcommand\unitWeight{w}

\newcommand\block{k}
\newcommand\Block{K}
\newcommand\blockweight{w}
\newcommand\blocklist{\mathcal{B}} %% Partition

\newcommand\Lookahead{L}
\newcommand\lookahead{\ell}
\newcommand\Lseq{\mathbb{L}}
\newcommand\Ninter{S} %% number of intermediate timepoints
\newcommand\ninter{s}
\newcommand\Nguide{K} %% number of lookahead particles
\newcommand\nguide{k}
\newcommand\lookaheadEnd{\min(\time+\Lookahead,\Time)}
\newcommand\guideFunc{g}
\newcommand\vmeasure{\mathrm{v}}
\newcommand\mmeasure{\mathrm{m}}
\newcommand\emeasure{\mathrm{e}}

\newcommand\artNoise{\epsilon}

\newcommand\iGIRF{IGIRF}
\newcommand\iEnKF{IEnKF}
\newcommand\iUBF{IUBF}

\newcommand*\colvec[3][]{
    \begin{pmatrix}\ifx\relax#1\relax\else#1\\\fi#2\\#3\end{pmatrix}
}

\newcommand\IF{A}  % adapted sample
\newcommand\IP{P}  %  adapted proposal
\newcommand\LCP{P}  % local prediction
\newcommand\LCF{F}  % local filter

\newcommand\Nit{M}
\newcommand\nit{m}
\newcommand\ntheta{K}

\newcommand\slot[1]{\code{#1}}
\newcommand\class[1]{class `\code{#1}'}
\newcommand\Class[1]{Class `\code{#1}'}

\newcommand\prob{\mathbb{P}}

\newcommand\given{{\,\vert\,}}

\newcommand\seq[2]{{#1}\!:\!{#2}}
\newcommand\mydot{{\,\cdot\,}}

\newcommand\giventh{{\hspace{0.5mm};\hspace{0.5mm}}}
\newcommand\normal{{\mathrm{Normal}}}
\newcommand\argequals{{\,=\,}}

\newcommand\bigO{\mathcal{O}}
\newcommand\loglik{\lambda}

\newcommand\loglikComponent{c}

\newcommand\param{\,;}
\newcommand\mycolon{{\hspace{0.6mm}:\hspace{0.6mm}}}

\newcommand\Var{\mathrm{Var}}
\newcommand\var{\Var}

\newcommand\dist{\mathrm{dist}}

\newcommand\gravity{g}
\newcommand\cohort{c}
\newcommand\popTimeAvg{\mathrm{pop}}
\newcommand\betaBar{\overline{\beta}}
\newcommand\measlesAmplitude{A}
\newcommand\measlesSeasonality{\mathrm{seas}}
\newcommand\measlesMeasurementOD{\tau}
\newcommand\measlesImmigration{\iota}
\newcommand\Rzero[1]{\mathcal{R}_{0#1}}

\author{Kidus Asfaw$^1$, Joonha Park$^2$, Aaron A. King$^1$ and Edward L. Ionides$^{1*}$\\
\vspace{2mm}
\small{$^1$University of Michigan, $^2$University of Kansas, $^*$correspondence to \texttt{ionides@umich.edu} } }

\title{A tutorial on spatiotemporal partially observed Markov process models via the R package spatPomp}

\date{}

\IfFileExists{upquote.sty}{\usepackage{upquote}}{}
\begin{document}

\maketitle

\begin{abstract}
We describe a computational framework for modeling and statistical inference on high-dimensional stochastic dynamic systems.
Our primary motivation is the investigation of metapopulation dynamics arising from a collection of spatially distributed, interacting biological populations.
To make progress on this goal, we embed it in a more general problem: inference for a collection of interacting partially observed nonlinear non-Gaussian stochastic processes.
Each process in the collection is called a unit;
in the case of spatiotemporal models, the units correspond to distinct spatial locations.
The dynamic state for each unit may be discrete or continuous, scalar or vector valued.
In metapopulation applications, the state can represent a structured population or the abundances of a collection of species at a single location.
We consider models where the collection of states has a Markov property.
A sequence of noisy measurements is made on each unit, resulting in a collection of time series.
A model of this form is called a spatiotemporal partially observed Markov process (SpatPOMP).
The \proglang{R} package \pkg{\spatPomp} provides an environment for implementing SpatPOMP models, analyzing data using existing methods, and developing new inference approaches.
Our presentation of \pkg{\spatPomp} reviews various methodologies in a unifying notational framework.
We demonstrate the package on a simple Gaussian system and on a nontrivial epidemiological model for measles transmission within and between cities.
We show how to construct user-specified SpatPOMP models within \pkg{\spatPomp}.

\vspace{3mm}

\noindent
This version was compiled on {\today} using \pkg{spatPomp} 0.34.2 with \pkg{pomp} 5.7.1.0 and \proglang{R} 4.3.3. Source code for this article is at \code{https://github.com/ionides/spatPomp-article}.
Materials are provided under the Creative Commons Attribution License. 

\end{abstract}

\section{Introduction}

A spatiotemporal partially observed Markov process (SpatPOMP) model consists of incomplete and noisy measurements of a latent Markov process having spatial as well as temporal structure.
A SpatPOMP model is a special case of a vector-valued partially observed Markov process (POMP) where the latent states and the measurements are indexed by a collection of spatial locations known as units.
Many biological, social and physical systems have the spatiotemporal structure, dynamic stochasticity and imperfect observability that characterize SpatPOMP models.
This paper discusses investigation of SpatPOMP models using the \pkg{spatPomp} software package \citep{R:spatPomp}, written in \proglang{R} \citep{R}.

Modeling and inference for spatiotemporal dynamics has long been considered a central challenge in ecology and epidemiology.
\citet{bjornstad01} identified six challenges of data analysis for ecological and epidemiological dynamics: (i) combining measurement noise and process noise; (ii) including covariates in mechanistically plausible ways; (iii) continuous time models; (iv) modeling and estimating interactions in coupled systems; (v) dealing with unobserved variables; (vi) spatiotemporal models.
Challenges (i) through (v) require nonlinear time series analysis methodology, and this has been successfully addressed over the past two decades via the framework of POMP models.
Software packages such as \pkg{pomp} \citep{king16}, \pkg{nimble} \citep{michaud21}, \pkg{LiBBi} \citep{murray15} and \pkg{mcstate} \citep{fitzjohn20} nowadays provide routine access to widely applicable modern inference algorithms for POMP models, as well as platforms for sharing models and data analysis workflows.
However, the Monte Carlo methods on which these packages depend do not scale well for high-dimensional systems and so are not practically applicable to SpatPOMP models.
Thus, challenge (vi) requires state-of-the-art algorithms with favorable scalability.

The \pkg{\spatPomp} package brings together general purpose methods for carrying out Monte Carlo statistical inference that meet all the requirements (i) through (vi).
For this purpose, \pkg{\spatPomp} provides an abstract representation for specifying SpatPOMP models.
This ensures that SpatPOMP models formulated with the package can be investigated using a range of methods, and that new methods can be readily tested on a range of models.
In its current form, \pkg{\spatPomp} is appropriate for data analysis with a moderate number of spatial units (say, 100) having nonlinear and non-Gaussian dynamics. 
In particular, \pkg{\spatPomp} is not targeted at very large spatiotemporal systems such as those that arise in geophysical data assimilation \citep{anderson09}.
Spatiotemporal systems with Gaussian dynamics can be investigated with \pkg{\spatPomp}, but a variety of alternative methods and software are available in this case \citep{wikle19,sigrist15,cappello20}.

The \pkg{\spatPomp} package builds on the \pkg{pomp} package described by \citet{king16}.
Mathematically, a SpatPOMP model is also a POMP model, and this property is reflected in the object-oriented design of \pkg{\spatPomp}.
The package is implemented using S4 classes \citep{chambers98,genolini08,wickham2019advanced} and the basic \class{spatPomp} extends the \class{pomp} provided by \pkg{pomp}.
This allows new methods to be checked against extensively tested methods in the low-dimensional settings for which POMP algorithms are effective.
However, standard Monte Carlo statistical inference methods for nonlinear POMP models suffer from a \emph{curse of dimensionality} \citep{bengtsson08}.
Extensions of these methods for situations with more than a few units must, therefore, take advantage of the special structure of SpatPOMP models.
Figure \ref{fig:usecase} illustrates the use case of the \pkg{\spatPomp} package relative to the \pkg{pomp} package and methods that use Gaussian approximations to target models with massive dimensionality.
Highly scalable methods, such as the Kalman filter and ensemble Kalman filter, entail approximations that may be inappropriate for nonlinear, non-Gaussian, count-valued models arising in metapopulation systems.

\begin{figure}[h]
  \centering
\begin{tikzpicture}
\begin{axis}[
axis lines=left,
xtick=\empty,
xmin=0,
xmax=7,
xlabel={Nonlinearity},
ytick=\empty,
ymin=0,
ymax=5.5,
ylabel={Dimension},
legend style={
at={(0,0)},
anchor=north east,at={(axis description cs:1,1)}}]
]
\addplot [forget plot,white,name path=A,domain=0:5] {5};
\addplot [forget plot,white,name path=B,domain=1:6] {1};
\addplot [forget plot,name path=C,domain=0:6] {0};

\addplot[pattern color=gray!90,pattern=north east lines] fill between [
of=A and C,
soft clip={domain=0:1},
];
\addlegendentry{KF}
\addplot[pattern color=gray!90,pattern=dots] fill between [
of=A and B,
soft clip={domain=1:2},
];
\addlegendentry{EnKF}

\addplot[pattern color=gray!90,pattern=grid] fill between [
of=B and C,
soft clip={domain=1:5.5},
];
\addlegendentry{PF}

%pomp
\draw (3,0.50) ellipse (2.5cm and 0.50cm);
\node[] at (axis cs: 6.025,0.5) {\texttt{pomp}};
%spatPomp
\draw (3,1.5) ellipse (2.5cm and 0.50cm);
\node[] at (axis cs: 5.65,2.0) {\texttt{spatPomp}};

\end{axis}
\end{tikzpicture}
\caption{The use case for the \pkg{\spatPomp} package. For statistical inference of models that are approximately linear and Gaussian, the Kalman Filter (KF) is an appropriate method. If the nonlinearity in the problem increases moderately but the dimension of the problem is very large (e.g. geophysical models), the ensemble Kalman Filter (EnKF) is useful. In low-dimensional but very nonlinear settings, the particle filter (PF) is widely applicable and the \pkg{pomp} package targets such problems. The \pkg{\spatPomp} package and the methods implemented in it are intended for statistical inference for nonlinear models that are of moderate dimension. The nonlinearity in these models (e.g. epidemiological models) is problematic for Gaussian approximations and the dimensionality is large enough to make the particle filter unstable.}
\label{fig:usecase}
\end{figure}

A SpatPOMP model is characterized by the transition density for the latent Markov process and unit-specific measurement densities.
Once these elements are specified, calculating and simulating from all joint and conditional densities are well defined operations.
However, different statistical methods vary in the operations they require.
Some methods require only simulation from the transition density whereas others require evaluation of this density.
Some methods avoid working with the model directly, replacing it by an approximation, such as a linearization.
For a given model, some operations may be considerably easier to implement and so it is useful to classify inference methods according to the operations on which they depend.
In particular, an algorithm is said to be \emph{plug-and-play} if it utilizes simulation of the latent process but not evaluation of transition densities \citep{breto09,he10}.
Simulators are relatively easy to implement for many SpatPOMP models, and so plug-and-play methodology facilitates the investigation of a variety of models that may be scientifically interesting but mathematically inconvenient.
Modern plug-and-play algorithms can provide statistically efficient likelihood-based or Bayesian inference.
The computational cost of plug-and-play methods may be considerable, due to the large number of simulations involved.
Nevertheless, the practical utility of plug-and-play methods for POMP models has been amply demonstrated in scientific applications.
In particular, plug-and-play methods implemented using \pkg{pomp} have facilitated various scientific investigations \citep[e.g.,][]{king08,bhadra11,shrestha11,shrestha13,earn12,roy13,blackwood13,blackwood13b,he13,breto14,blake14,martinez-bakker15,bakker16,becker16,buhnerkempe17,ranjeva17,marino19,pons-salort18,becker19,kain20,stocks20}.
The \pkg{spatPomp} package has been used to develop and demonstrate plug-and-play methodology for SpatPOMP models \citep{ionides22,ionides23,ning23-ibpf}.
Scientific applications are starting to emerge \citep{zhang22,wheeler24,li24}.

The remainder of this paper is organized as follows.
\Cref{sec:background} defines mathematical notation for SpatPOMP models and relates this to their representation as objects of \class{\spatPomp} in the \pkg{\spatPomp} package.
\Cref{sec:filtering} introduces likelihood evaluation via several spatiotemporal filtering methods.
\Cref{sec:inference} describes parameter estimation algorithms which build upon these filtering techniques.
\Cref{sec:brownian} constructs a simple linear Gaussian SpatPOMP model and uses this example to illustrate statistical inference.
\Cref{sec:measles} presents the construction of spatially structured compartment models for population dynamics, in the context of coupled measles dynamics in UK cities;
this demonstrates the kind of nonlinear stochastic system primarily motivating the development of \pkg{\spatPomp}.
\cref{sec:conclusion} is a concluding discussion.

\section{SpatPOMP models and their representation in spatPomp}
\label{sec:background}

\begin{figure}[h]
  \centering
\begin{tikzpicture}
[
  main/.style = {draw, circle}, % circles
  ssp/.style={draw, rounded corners,inner sep=2pt} % state space
] 
		\node[main] (X10) {$X_{1,0}$}; 
		\node[main,draw=white,text=black] (D0) [right of=X10] {$\cdots$}; 
		\node[main] (XU0) [right of=D0] {$X_{U,0}$}; 
		\node[main] (X11) [right=1cm of XU0] {$X_{1,1}$}; 
		\node[main,draw=white,text=black] (XD1) [right of=X11] {$\cdots$}; 
		\node[main] (XU1) [right of=XD1] {$X_{U,1}$};
		\node[ssp, fit=(X10) (D0) (XU0)](X0){};
		\draw[decoration={brace,raise=3pt},decorate](X0.north west) -- node[above=6pt] {$\vec{X}_0$} (X0.north east);
		\node[ssp, fit=(X11) (XD1) (XU1)](X1){};
		\draw[decoration={brace,raise=3pt},decorate](X1.north west) -- node[above=6pt] {$\vec{X}_1$} (X1.north east);
		\draw[->] (X0) -- (X1);
		\node[main] (Y11) [below=1.5cm of X11] {$Y_{1,1}$}; 
		\node[main,draw=white,text=black] (YD1) [right of=Y11] {$\cdots$}; 
		\node[main] (YU1) [below=1.5cm of XU1] {$Y_{U,1}$};
		\node[ssp, fit=(Y11) (YD1) (YU1)](Y1){};
		\draw[decoration={brace,mirror,raise=3pt},decorate](Y1.south west) -- node[below=6pt] {$\vec{Y}_1$} (Y1.south east);
		\draw[->] (X11.south) -- (Y11.north) node[midway, right=0.6cm]{$\cdots$};
		\draw[->] (XU1.south) -- (YU1.north);
    \draw [->] (X1.east) -- ++(\FixedLengthArrow) node[right] (midstates) {$\cdots$};
    \node[main,inner sep=2.5pt] (X1N) [right=1cm of midstates] {$X_{1,N}$}; 
		\node[main,draw=white,text=black] (XDN) [right of=X1N] {$\cdots$}; 
		\node[main,inner sep=2.5pt] (XUN) [right of=XDN] {$X_{U,N}$};
		\node[ssp, fit=(X1N) (XDN) (XUN)](XN){};
		\draw[decoration={brace,raise=3pt},decorate](XN.north west) -- node[above=6pt] {$\vec{X}_N$} (XN.north east);
		\draw [->] (midstates) -- (XN);
		\node[main,inner sep=2.8pt] (Y1N) [below=1.5cm of X1N] {$Y_{1,N}$}; 
		\node[main,draw=white,text=black] (YDN) [right of=Y1N] {$\cdots$}; 
		\node[main,inner sep=2.8pt] (YUN) [below=1.5cm of XUN] {$Y_{U,N}$};
		\node[ssp, fit=(Y1N) (YDN) (YUN)](YN){};
		\draw[decoration={brace,mirror,raise=3pt},decorate](YN.south west) -- node[below=6pt] {$\vec{Y}_N$} (YN.south east);
		\draw[->] (X1N.south) -- (Y1N.north) node[midway, right=0.6cm]{$\cdots$};
		\draw[->] (XUN.south) -- (YUN.north);
		\node (YD) [right=1cm of Y1] {$\cdots$};
	\end{tikzpicture}
	\caption{Conditional dependence diagram for a spatiotemporal partially observed Markov process (SpatPOMP) model. The latent dynamic process is $\{\vec{X}(t), t_0\le t\le t_{\Time}\}$. At observation times $t_n$, the value of the latent process is denoted by $\vec{X}_{n}=\big(X_{1,n},\dots,X_{U,n}\big)$. The partial and noisy observations at this times are modeled by $\vec{Y}_{n}=\big(Y_{1,n},\dots,Y_{U,n}\big)$.} \label{fig:spatpomp}
\end{figure}

We set up notation for SpatPOMP models extending the POMP notation of \citet{king16}.
A diagrammatic representation is given in Figure \ref{fig:spatpomp}.
Suppose there are $\Unit$ units labeled $1\mycolon\Unit=\{1,2,\dots,\Unit\}$.
Let $t_1<t_2<\dots <t_\Time$ be a collection of times at which measurements are recorded on one or more units, and let $t_0$ be some time preceding $t_1$ at which we initialize our model.
We observe a measurement $\data{y}_{u,n}$ on unit $u$ at time $t_n$, where $\data{y}_{u,n}$ could take the value \code{NA} if no measurement was recorded. 
We postulate a latent stochastic process taking value $\vec{X}_n=(X_{1,n},\dots,X_{U,n})$ at time $t_n$, with boldface denoting a collection of random variables across units.
The observation $\data{y}_{u,n}$ is modeled as a realization of an observable random variable $Y_{u,n}$, and we suppose that the collection of observable random variables are conditionally independent given the collection of latent random variables.
The process $\vec{X}_{0:N}=(\vec{X}_0,\vec{X}_1,\dots,\vec{X}_N)$ is required to have the Markov property, i.e., $\vec{X}_{0:n-1}$ and $\vec{X}_{n+1:N}$ are conditionally independent given $\vec{X}_n$.
Optionally, there may be a continuous time process $\vec{X}(t)$ defined for $t_0\le t\le t_N$ such that $\vec{X}_n=\vec{X}(t_n)$.

Let $f_{\vec{X}_{0:\Time},\vec{Y}_{1:\Time}}(\vec{x}_{0:\Time},\vec{y}_{1:\Time};\theta)$  be the joint density of $X_{1:\Unit,0:\Time}$ and $Y_{1:\Unit,1:\Time}$ evaluated at  $x_{1:\Unit,0:\Time}$ and $y_{1:\Unit,1:\Time}$, depending on
an unknown parameter vector, $\theta$.
We do not distinguish between continuous and discrete spaces for the latent and observation processes, so the term \emph{density} encompasses probability mass functions.
The SpatPOMP structure permits a factorization of the joint density in terms of the initial density,
$f_{\vec{X}_{0}}(\vec{x}_{0};\vectheta)$,
the transition density,
$f_{\vec{X}_{\time}|\vec{X}_{\time-1}}(\vec{x}_{\time}\given \vec{x}_{\time-1}\giventh\vectheta)$,
and the unit measurement density,
$f_{Y_{\unit,\time} | X_{\unit,\time}} (y_{\unit,\time} | x_{\unit,\time}\giventh \vectheta)$, given by
\begin{equation}
\nonumber
  f_{\vec{X}_{0:\Time},\vec{Y}_{1:\Time}}(\vec{x}_{0:\Time},\vec{y}_{1:\Time};\vectheta)
  =
  f_{\vec{X}_0}(\vec{x}_0;\vectheta)\,
  \prod_{\time=1}^{\Time}\!
    f_{\vec{X}_{\time} | \vec{X}_{\time-1}}(\vec{x}_{\time}|\vec{x}_{\time-1};\vectheta)\,
    \prod_{\unit=1}^{\Unit}f_{Y_{\unit,\time} | X_{\unit,\time}} (y_{\unit,\time} | x_{\unit,\time}\giventh \vectheta).
\end{equation}
This notation allows $f_{\vec{X}_{\time}|\vec{X}_{\time-1}}$ and $f_{Y_{\unit,\time} | X_{\unit,\time}}$ to depend on $\time$ and $\unit$, thereby permitting models for temporally and spatially inhomogeneous systems.

\subsection{Implementation of SpatPOMP models}
\label{sec:implementation}

A SpatPOMP model is represented in \pkg{\spatPomp} by an object of \class{\spatPomp}.
Slots in this object encode the components of the SpatPOMP model, and can be filled or changed using the constructor function \code{\spatPomp()} and various other convenience functions.
Methods for the \class{\spatPomp} (i.e., functions defined in the package which take a \class{\spatPomp} object as their first argument) use these components to carry out computations on the model.
\Cref{tab:notation} lists elementary methods for a \class{\spatPomp} object, and their translations into mathematical notation.

\begin{table}[t!]
   \begin{center}
     \begin{tabular}{llll}
       \hline
       Method &Argument to &Mathematical terminology \\
       & \code{\spatPomp()} & \\
       \hline
      \code{dunit\_measure} &\code{dunit\_measure} & Evaluate $f_{Y_{\unit,\time}|X_{\unit,\time}}( y_{\unit,\time} \given x_{\unit,\time}\giventh \vectheta)$\\
      \code{runit\_measure} &\code{runit\_measure} & Simulate from $f_{Y_{\unit,\time}|X_{\unit,\time}}( y_{\unit,\time} \given x_{\unit,\time}\giventh \vectheta)$\\
      \code{eunit\_measure} &\code{eunit\_measure} & Evaluate $\emeasure_{\unit,\time}(x,\vectheta)=\E[Y_{\unit,\time}\given X_{\unit,\time}=x\giventh\vectheta]$\\
      \code{vunit\_measure} &\code{vunit\_measure} & Evaluate $\vmeasure_{\unit,\time}(x,\vectheta)=\Var[Y_{\unit,\time}\given X_{\unit,\time}=x\giventh\vectheta]$\\
      \code{munit\_measure} &\code{munit\_measure} & $\mmeasure_{\unit,\time}(x,V,\vectheta)=\vec{\psi}$ if $\vmeasure_{\unit,\time}(x,\vec{\psi})=V$, $\emeasure_{\unit,\time}(x,\vec{\psi})=\emeasure_{\unit,\time}(x,\vectheta)$\\
       \code{rprocess} &\code{rprocess} &Simulate from $f_{\vec{X}_{\time}|\vec{X}_{\time-1}}( \vec{x}_{\time} \given \vec{x}_{\time-1}\giventh \vectheta)$\\
       \code{dprocess} &\code{dprocess} &Evaluate $f_{\vec{X}_{\time}|\vec{X}_{\time-1}}( \vec{x}_{\time} \given \vec{x}_{\time-1}\giventh \vectheta)$\\
       \code{rmeasure} &\code{rmeasure} &Simulate from $f_{\vec{Y}_{\time}|\vec{X}_{\time}}( \vec{y}_{\time} \given \vec{x}_{\time}\giventh \vectheta)$\\
       \code{dmeasure} &\code{dmeasure} &Evaluate $f_{\vec{Y}_{\time}|\vec{X}_{\time}}( \vec{y}_{\time} \given \vec{x}_{\time}\giventh \vectheta)$\\
       \code{rprior} &\code{rprior} &Simulate from the prior distribution $\pi(\vectheta)$\\
       \code{dprior} &\code{dprior} &Evaluate the prior density $\pi(\vectheta)$\\
       \code{rinit} &\code{rinit} &Simulate from $f_{\vec{X}_0}( \vec{x}_0 \giventh \vectheta)$\\
       \code{timezero} &\code{t0} &$t_0$\\
       \code{time} &\code{times} &$t_{1:\Time}$\\
       \code{obs} &\code{data} &$\data{\vec{y}}_{1:\Time}$\\
       \code{states} & --- &$\vec{x}_{0:\Time}$\\
       \code{coef} &\code{params} &$\vectheta$\\
       \hline
    \end{tabular}
  \end{center}
\caption{
Elementary methods for \class{\spatPomp} objects, the argument used to assign them via the \code{\spatPomp} constructor function, and their definition in mathematical notation.
}
\label{tab:notation}
\end{table}

\Class{spatPomp} inherits from the \class{pomp} defined by the \pkg{pomp} package.
In particular, \pkg{\spatPomp} extends \pkg{pomp} by the addition of unit-level specification of the measurement model.
This reflects the modeling assumption that measurements are carried out independently in both space and time, conditional on the value of the spatiotemporal latent process.
There are five unit-level functionalities of \class{spatPomp} objects:
\code{dunit\_measure}, \code{runit\_measure}, \code{eunit\_measure},
\code{vunit\_measure} and \code{munit\_measure}.
These model components are specified by the user via an argument to the \code{spatPomp()} constructor function of the same name.

All the model components of a \class{spatPomp} object are listed in \cref{tab:notation}.
It is not necessary to supply every component---only those that are required to run an algorithm of interest.
For example, the functions \code{eunit\_measure} and \code{vunit\_measure}, calculating the expectation and variance of the measurement model, are used by the ensemble Kalman filter (EnKF, \cref{sec:enkf}) and iterated EnKF (\cref{sec:ienkf}).
The function \code{munit\_measure} returns a parameter vector corresponding to given mean and variance, used by one of the options for a guided intermediate resampling filter (GIRF, \cref{sec:girf}) and iterated GIRF (\cref{sec:igirf}).

\subsection{Examples included in the package}
\label{section:included}
Though users can construct arbitrary \class{spatPomp} models, pre-built examples are available via the functions \code{bm()}, \code{bm2()}, \code{gbm()}, \code{he10()}, \code{lorenz()}, and \code{measles()}.
These create \class{spatPomp} models with user-specified dimensions for correlated Brownian motion models (\code{bm}, \code{bm2}, \code{gbm}), the Lorenz-96 atmospheric model of \citep{lorenz96} (\code{lorenz}), and spatiotemporal susceptible-exposed-infected-recovered epidemiological models (\code{he10}, \code{measles}).
Users may find the source code for these examples useful as templates for the construction of custom models.
In Section \ref{sec:measles}, we work through the construction of a scientifically motivated \class{spatPomp} object.

\begin{knitrout}
\definecolor{shadecolor}{rgb}{1, 1, 1}\color{fgcolor}\begin{figure}

\includegraphics[width=1\linewidth]{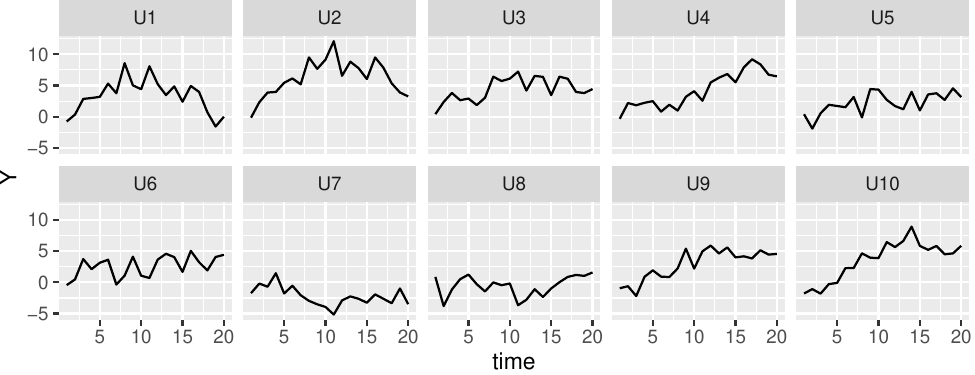} \hfill{}

\caption[Result of executing \code{plot(bm10)}, where \code{bm10} is the \class{spatPomp} object representing a simulation from a 10-dimensional correlated Brownian motions model with 20 observations that are one unit time apart (see text)]{Result of executing \code{plot(bm10)}, where \code{bm10} is the \class{spatPomp} object representing a simulation from a 10-dimensional correlated Brownian motions model with 20 observations that are one unit time apart (see text).}\label{fig:bm10-simplot}
\end{figure}

\end{knitrout}

Our first \code{spatPomp} example model is a simulation of $U=10$ correlated Brownian motions each with $N=20$ measurements, constructed by executing \code{bm10 <- bm(U = 10, N = 20)}.
The correlation structure and other model details are discussed in Section \ref{sec:brownian}.
We can view the data using \code{plot(bm10)}, shown in \cref{fig:bm10-simplot}.
For customized plots using the many plotting options in \proglang{R} for \class{data.frame} objects, the data in \code{bm10} can be extracted using \code{as.data.frame(bm10)}.
The accessor functions in \cref{tab:notation} extract various components of \code{bm10} via \code{timezero(bm10)}, \code{time(bm10)}, \code{obs(bm10)}, \code{states(bm10)}, \code{coef(bm10)}.
The internal representation of all the components of the object can be inspected via \code{spy(bm10)}.

\subsection{Data and observation times}

The only mandatory arguments to the \code{spatPomp()} constructor are \code{data}, \code{times}, \code{units} and \code{t0}.
The \code{data} argument requests a \class{data.frame} object containing observations for each spatial unit at each observation time.
Missing data for some or all units at each observation time can be coded as \code{NA}.
It is the user's responsibility to specify a measurement model that assigns an appropriate probability to the value \code{NA}.
The name of the \code{data} column containing observation times is supplied to the \code{times} argument; the name of the column containing the unit names is supplied to the \code{units} argument.
The \code{t0} argument supplies the initial time from which the dynamic system is modeled, which should be no greater than the first observation time.

We may also wish to add parameter values, latent state values, and some or all of the model components from \cref{tab:notation}.
We need to define only those components necessary for operations we wish to carry out.
In particular, plug-and-play methodology by definition never uses \code{dprocess}. An empty \slot{dprocess} slot in a \class{\spatPomp} object is therefore acceptable unless a non-plug-and-play algorithm is attempted.

\subsection{Initial conditions}
\label{subsec:init}

The initial state of the latent process, $\vec{X}_0=\vec{X}(t_0;\vectheta)$, is a draw from the initial distribution, $f_{\vec{X}_0}(\vec{x}_0\giventh\vectheta)$. 
If the initial conditions are known, there is no dependence on $\vectheta$.
Alternatively, there may be components of the $\vectheta$ having the sole function of specifying $\vec{X}_0$.
These components are called \emph{initial value parameters} (IVPs).
By contrast, parameters involved in the transition density or measurement density are called \emph{regular parameters} (RPs).
This gives rise to a decomposition of the parameter vector, $\theta=(\theta_{\RP},\theta_{\IVP})$.
We may specify $f_{\vec{X}_0}(\vec{x}_0\giventh \theta_{\RP},\theta_{\IVP})$ to be a point mass at $\theta_{\IVP}$, in which case $\theta_{\IVP}$ exactly corresponds to $\vec{X}_0$.
The \code{bm10} model has this structure, and the initialization can be tested by \code{rinit(bm10)}.
The measles model of \cref{sec:measles:spatPomp} specifies $\vec{X}_0$ as a deterministic function of $\theta_{\IVP}$, but not an identity map since it is convenient to describe latent states as counts and the corresponding IVPs as proportions. 

\subsection{Parameters}
\label{sec:params}

Many \pkg{spatPomp} methods require a named numeric vector to represent a parameter, $\theta$.
In addition to the initial value parameters introduced in \cref{subsec:init}, a parameter can be \emph{unit-specific} or \emph{shared}.
A unit-specific parameter has a distinct value defined for each unit, and a shared parameter is one without that structure.
We can write $\theta=(\phi,\psi_{1:U})$, where $\phi$ is the vector of shared parameters and $\psi_{\unit}$ is the vector of unit-specific parameters for unit $\unit$.
The unit methods in \Cref{tab:notation} require only $\phi$ and $\psi_{\unit}$ when evaluated on unit $u$.
A shared/unit-specific structure can be combined with an RP/IVP decomposition to give
\begin{equation}
\nonumber
\theta=\big(\phi^{}_{\RP},\phi^{}_{\IVP},\psi^{}_{\RP,1:U},\psi^{}_{\IVP,1:U}\big).
\end{equation}
The \code{bm10} and measles examples are coded with unit-specific IVPs and shared RPs.
The dimension of the parameter space can increase quickly with the number of unit-specific parameters.
Shared parameters provide a more parsimonious description of the system, which is desirable when it is consistent with the  data.

\subsection{Covariates}
\label{subsec:covariates}

Scientifically, one may be interested in the impact of a vector-valued covariate process, $\mathbf{Z}(t)$, on the latent dynamic system.
Our modeling framework allows the transition density,
$f_{\vec{X}_{\time}|\vec{X}_{\time-1}}$, and the measurement density,
$f_{\vec{Y}_{\time}|\vec{X}_{\time}}$, to depend arbitrarily on time, and this includes the possibility of dependence on one or more covariates.
A covariate process is called {\it shared} if, at each time, it takes single value which influences all the units.
A {\it unit-specific} covariate process, $\mathbf{Z}(t)=Z_{1:U}(t)$, has a value, $Z_{\unit}(t)$, for each unit, $u$.
In \pkg{\spatPomp}, covariate processes can be supplied as a \class{data.frame} object to the \code{covar} argument of the \code{spatPomp()} constructor function.
This \code{data.frame} requires a column for time, spatial unit, and each of the covariates.
If any of the variables in the covariates \code{data.frame} is common among all units the user must supply the variable names as \class{character} vectors to the \code{shared\_covarnames} argument of the \code{spatPomp()} constructor function.
All covariates not declared as shared are assumed to be unit-specific.
\pkg{spatPomp} manages the task of presenting interpolated values of the covariates to the elementary model functions at the time they are called.
An example implementing a SpatPOMP model with covariates is presented in \cref{sec:measles}.

\subsection{Specifying model components using C snippets}
\label{sec:csnippets}

The \pkg{spatPomp} function \code{spatPomp\_Csnippet} extends the Csnippet facility in \pkg{pomp} which allows users to specify the model components in \Cref{tab:notation} via fragments of \proglang{C} code.
The use of Csnippets permits computationally expensive calculations to take advantage of the performance of \proglang{C}.
The Csnippets are compiled in a suitable environment by a call to \code{spatPomp()}, however, \code{spatPomp()} needs some help to determine which variables should be defined.
In behavior inherited from \pkg{pomp}, the names of the parameters and latent variables must be supplied to \code{spatPomp} using the \code{paramnames} and \code{unit\_statenames} arguments, and the names of observed variables and covariates are extracted from the supplied data.
In \pkg{spatPomp}, unit-specific variable names can be supplied as needed via arguments to \code{spatPomp\_Csnippet}.
These can be used to specify the five \code{unit\_measure} model components in \cref{tab:notation} which specify properties of the spatially structured measurement model characteristic of a SpatPOMP.
For a \code{unit\_measure} Csnippet, automatically defined variables also include the number of units, \code{U}, and an integer \code{u} corresponding to a numeric unit from \code{0} to \code{U-1}.

A Csnippet can look similar to a domain-specific language.
For example, the unit measurement density for the \code{bm10} example is simply
\begin{knitrout}
\definecolor{shadecolor}{rgb}{1, 1, 1}\color{fgcolor}\begin{kframe}
\begin{verbatim}
R> spatPomp_Csnippet("lik = dnorm(Y,X,tau,give_log);")
\end{verbatim}
\end{kframe}
\end{knitrout}
Here, \code{spatPomp} makes all the required variables available to the Csnippet: the unit state name variable, \code{X}; the unit measurement variable, \code{Y}; the parameter, \code{tau}; and a logical flag \code{give\_log} indicating whether the desired output is on log scale, following a standard convention for the \proglang{C} interface to \proglang{R} distribution functions \citep{R:extensions}.
For models of increasing complexity the full potential of the \proglang{C} language is available.
In particular, additional \proglang{C} variables can be defined when needed, as demonstrated in \cref{sec:measles}.

Unlike the strict unit structure required for the measurement process, the latent process for a SpatPOMP model can have arbitrary spatial dependence between units.
We cannot in general define the full coupled dynamics by a collection of \code{runit\_process} functions defined separately for each unit.
Therefore, \pkg{spatPomp} relies on a \code{rprocess} function defined exactly as for \pkg{pomp}.
A \pkg{spatPomp} Csnippet for \code{rprocess} will typically involve a computation looping through the units, which requires access to location data used to specify the interaction between units.
The location data can be made available to the Csnippet using the \code{globals} argument.
Further details on this are postponed to \cref{sec:measles}.

\subsection{Simulation}
\label{subsec:simulation}

A first step to explore a SpatPOMP model is to simulate stochastic realizations of the latent process and the resulting measurements.
This is carried out by \code{simulate()} which requires specification of \code{rprocess} and \code{rmeasure}.
For example, \code{simulate(bm10)} produces a new object of \class{\spatPomp} for which the original data have been replaced with a simulation from the specified model.
Unless a \code{params} argument is supplied, the simulation will be carried out using the parameter vector in \code{coef(bm10)}.
Optionally, \code{simulate} can be made to return a \class{data.frame} object by supplying the argument \code{format=`data.frame'} in the call to \code{simulate}.

\section{Likelihood evaluation}
\label{sec:filtering}

We describe algorithms for likelihood evaluation in this section, followed by algorithms for likelihood maximization in \cref{sec:inference}.
These tools are subsequently demonstrated in \cref{sec:brownian}.

Likelihood evaluation for SpatPOMP models is effected via a filtering calculation.
The curse of dimensionality associated with spatiotemporal models can make filtering for SpatPOMP models computationally challenging, even though a single likelihood evaluation cannot be more than a small step toward a complete likelihood-based inference workflow.
A widely used time-series filtering technique is the basic particle filter (PF) available as \code{pfilter} in the \pkg{pomp} package.
However, PF and many of its variations scale poorly with dimension \citep{bengtsson08,snyder15}.
Thus, in the spatiotemporal context, successful particle filtering requires state-of-the-art algorithms.
Below, we introduce four such algorithms implemented in the \pkg{\spatPomp} package: a guided intermediate resampling filter (GIRF) implemented as \code{girf}, an adapted bagged filter (ABF) implemented as \code{abf}, an ensemble Kalman filter (EnKF) implemented as \code{enkf}, and a block particle filter (BPF) implemented as \code{bpfilter}.

The filtering problem can be decomposed into two steps, prediction and filtering.
For all the filters we consider here, the prediction step involves simulating from the latent process model.
The algorithms differ primarily in their approaches to the filtering step, also known as the data assimilation step or the analysis step.
For PF, the filtering step is a weighted resampling from the prediction particles, and the instability of these weights in high dimensions is the fundamental scalability issue with the algorithm.
GIRF carries out this resampling at many intermediate timepoints with the goal of breaking an intractable resampling problem into a sequence of tractable ones.
EnKF estimates variances and covariances of the prediction simulations, and carries out an update rule that would be exact for a Gaussian system.
BPF carries out the resampling independently over a partition of the units, aiming for an inexact but numerically tractable approximation. 
ABF combines together many high-variance filters using local weights to beat the curse of dimensionality.
We proceed to describe these algorithms in more detail.

\subsection[GIRF]{The guided intermediate resampling filter (GIRF)}
\label{sec:girf}

The guided intermediate resampling filter \citep[GIRF,][]{park20} is an extension of the auxiliary particle filter \citep[APF,][]{pitt99}.
GIRF is appropriate for moderately high-dimensional SpatPOMP models with a continuous-time latent process.
All particle filters compute importance weights for proposed particles and carry out resampling to focus computational effort on particles consistent with the data \citep[see reviews by][]{arulampalam02,doucet11,kantas15}.
In the context of \pkg{pomp}, the \code{pfilter} function is discussed by \citet{king16}.
GIRF combines two techniques for improved scaling of particle filters: the use of a guide function and intermediate resampling.

The guide function steers particles using importance weights that anticipate upcoming observations.
Future measurements are considered up to a lookahead horizon, $\Lookahead$.
APF corresponds to a lookahead horizon $\Lookahead=2$, and a basic particle filter has $\Lookahead=1$.
Values $\Lookahead \le 3$ are typical for GIRF.

Intermediate resampling breaks each observation interval into $\Ninter$ sub-intervals, and carries out reweighting and resampling on each sub-interval.
Perhaps surprisingly, intermediate resampling can facilitate some otherwise intractable importance sampling problems \citep{delmoral15}.
APF and the basic particle filter correspond to $\Ninter=1$, whereas choosing $\Ninter=\Unit$ gives favorable scaling properties \citep{park20}.

\begin{algorithm}[p]
  \caption{\code{girf(P,Np{\argequals}$\Np\!$,
     Ninter{\argequals}$\Ninter\!$,
     Nguide{\argequals}$\Nguide\!$,
     Lookahead{\argequals}$\Lookahead$)},
     using notation from \cref{tab:notation} where \code{P} is a
     `\code{\spatPomp}' object equipped with
     \code{rprocess},
     \code{dunit\_measure},
     \code{rinit},
     \code{skeleton},
     \code{obs},
     \code{coef}.
  }
  \label{alg:girf}
  \KwIn{
   simulator for $f_{\vec{X}_{\time}|\vec{X}_{\time-1}}(\vec{x}_{\time}\given \vec{x}_{\time-1}\giventh\vectheta)$ and  $f_{\vec{X}_0}(\vec{x}_0\giventh\vectheta)$;
   evaluator for $f_{{Y}_{\unit,\time}|{X}_{\unit,\time}}({y}_{\unit,\time}\given {x}_{\unit,\time}\giventh\vectheta)$,
   and  $\vec{\mu}(\vec{x},s,t\giventh\vectheta)$;
   data, $\data{\vec{y}}_{1:\Time}$;
   parameter, $\vectheta$;
    number of particles, $\Np$;
    number of guide simulations, $\Nguide$;
    number of intermediate timesteps, $\Ninter$;
    number of lookahead lags, $\Lookahead$.
  }
  initialize:
  simulate $\vec{X}_{0,0}^{F,\np}\sim {f}_{\vec{X}_{0}}(\mydot\giventh {\vectheta})$ and set $\guideFunc^{F,\np}_{0,0}=1$  for $\np$ in $\seq{1}{\Np}$
  \;
  \For{$\time \,\, \mathrm{in} \,\, \seq{0}{\Time-1}$}{
    sequence of guide forecast times, $\Lseq=\seq{(\time+1)}{\lookaheadEnd}$
    \;
    guide simulations,
      $\vec{X}_{\Lseq}^{G,\np,\nguide}
      \sim
      {f}_{\vec{X}_{\Lseq}|\vec{X}_{\time}}
      \big( \mydot|\vec{X}_{\time,0}^{F,\np} \giventh {\vectheta} \big)$
 for $\np$ in $\seq{1}{\Np}$,  $\nguide$ in $\seq{1}{\Nguide}$
      \nllabel{alg:girf:guide:sim}\;
      guide residuals, $\vec{\epsilon}^{\np,\nguide}_{0,\lookahead}=
        \vec{X}_{\lookahead}^{G,\np,\nguide} -
	\vec{\mu}\big(
	  \vec{X}^{F,\np}_{\time},t_{\time},t_{\lookahead} \giventh{\vectheta}
	\big)$
	for $\np$ in $\seq{1}{\Np}$,  $\nguide$ in $\seq{1}{\Nguide}$, $\lookahead$ in $\Lseq$
      \nllabel{alg:girf:guide:resid}\;
    \For{$\ninter  \,\, \mathrm{in} \,\, \seq{1}{\Ninter}$}{
    \nllabel{alg:girf:loop:ninter}
      prediction simulations,
        ${\vec{X}}_{\time,\ninter}^{P,\np}
          \sim {f}_{{\vec{X}_{\time,\ninter}}|{\vec{X}_{\time,\ninter-1}}}
          \big(\mydot|{\vec{X}^{F,\np}_{\time,\ninter-1}} \giventh
	    {\vectheta}\big)$
      for $\np$ in $\seq{1}{\Np}$
          \nllabel{alg:girf:predictions}\;
      deterministic trajectory,
        $\vec{\mu}^{P,\np}_{\time,\ninter,\lookahead}
           = \vec{\mu}\big(
	     \vec{X}^{P,\np}_{\time,\ninter},t_{\time,\ninter},t_{\lookahead}
	     \giventh \vectheta \big)$
        for $\np$ in $\seq{1}{\Np}$, $\lookahead$ in $\Lseq$
        \nllabel{alg:girf:guide:skeleton}\;
      pseudo guide simulations,
        $\hat{\vec{X}}^{\np,\nguide}_{\time,\ninter,\lookahead} =
        \vec{\mu}^{P,\np}_{\time,\ninter,\lookahead} +
        \vec{\epsilon}^{\np,\nguide}_{\ninter-1,\lookahead} -
        \vec{\epsilon}^{\np,\nguide}_{\ninter-1,\time+1} +
        {\textstyle \sqrt{
            \frac{t_{\time+1}-t_{\time,\ninter}}{t_{\time+1}-t_{\time,0}}}
        }	\,
        \vec{\epsilon}^{\np,\nguide}_{\ninter-1,\time+1}$
	for $\np$ in $\seq{1}{\Np}$,  $\nguide$ in $\seq{1}{\Nguide}$, $\lookahead$ in $\Lseq$\;
      discount factor,
        $\eta_{\time, \ninter,\lookahead}
          = 1-(t_{\time+\lookahead}-t_{\time,\ninter})/\{(t_{\time+\lookahead}-t_{\max(\time+\lookahead-\Lookahead,0)})\cdot(1+\mathbbm{1}_{\Lookahead=1})\}$\;
% guide function,
        $ \displaystyle
\guideFunc^{P,\np}_{\time,\ninter}=
          \prod_{\lookahead \, \mathrm{in} \, \Lseq}
	  \,
          \prod_{\unit=1}^{\Unit}
          \left[
          \frac{1}{\Nguide}
	  \sum_{\nguide=1}^{\Nguide}
          f_{Y_{\unit,\lookahead}|X_{\unit,\lookahead}}
          \Big(
            \data{y}_{\unit,\lookahead}\given \hat{X}^{\np,\nguide}_{\unit,\time,\ninter,\lookahead}
	     \giventh \vectheta
          \Big)
          \right]^{\eta_{\time, \ninter,\lookahead}}$
        for $\np$ in $\seq{1}{\Np}$
        \nllabel{alg:girf:guideFunc}\\
%        weights
	for $\np$ in $\seq{1}{\Np}$,
        $w^{\np}_{\time,\ninter}=\left\{ \hspace{-1mm}
          \begin{array}{ll}
          f_{\vec{Y}_{\time}|\vec{X}_{\time}} \big(
            \vec{y}_{\time}\given \vec{X}^{F,\np}_{\time,\ninter-1}\giventh\vectheta
            \big) \,\,
         \guideFunc^{P,\np}_{\time,\ninter}
         \Big/ \guideFunc^{F,\np}_{\time,\ninter-1}
          & \mbox{if $\ninter=1$ and $\time\neq 0$} \\
            \guideFunc^{P,\np}_{\time,\ninter}
             \Big/ \guideFunc^{F,\np}_{\time,\ninter-1}
          & \mbox{else}
          \end{array} \right.$
        \nllabel{alg:girf:weights}\;
      log-likelihood component,
        $\loglikComponent_{\time,\ninter}=
          \log\Big(\Np^{-1}\,\sum_{\altNp=1}^{\Np}\!w^{\altNp}_{\time,\ninter}\Big)$\;
      normalized weights,
        $\tilde{w}^{\np}_{\time,\ninter}= w^{\np}_{\time,\ninter}\Big/\sum_{\altNp=1}^{\Np}w^{\altNp}_{\time,\ninter}$
	for $\np$ in $\seq{1}{\Np}$
	\;
        select resample indices,
        $\resampleIndex_{1:\Np}$
        with
        $\prob\left[\resampleIndex_{\np}=\altNp\right] =\tilde{w}^{\altNp}_{\time,\ninter}$
	for $\np$ in $\seq{1}{\Np}$
        \nllabel{alg:girf:systematic}\;
%      resample:
        $\vec{X}_{\time,\ninter}^{F,\np}=\vec{X}_{\time,\ninter}^{P,\resampleIndex_{\np}}\,$,
        $\; \guideFunc^{F,\np}_{\time,\ninter}= \guideFunc^{P,\resampleIndex_{\np}}_{\time,\ninter}\,$,
        $\; \vec{\epsilon}^{\np,\nguide}_{\ninter,\lookahead}=
	  \vec{\epsilon}^{\resampleIndex_{\np},\nguide}_{\ninter-1,\lookahead}$
        for $\np$ in $\seq{1}{\Np}$,  $\nguide$ in $\seq{1}{\Nguide}$, $\lookahead$ in $\Lseq$
        \nllabel{alg:girf:resample}\;
    }
  set $\vec{X}^{F,\np}_{\time+1,0}=\vec{X}^{F,\np}_{\time,\Ninter}$ and  $g^F_{\time+1,0,\np}=g^F_{\time,\Ninter,\np}$
  for $\np$ in $\seq{1}{\Np}$
  \;
  }
  \KwOut{
    log-likelihood,
      $\loglik^{\mbox{\tiny{GIRF}}}=
        \sum_{\time=0}^{\Time-1} \sum_{\ninter=1}^{\Ninter}
      \loglikComponent_{\time,\ninter}$, and
    filter particles, $\vec{X}^{F,1:\Np}_{\Time,0}$
  }
  \KwCplx{$\bigO\big({\Np\Lookahead\Unit\Time(\Nguide+\Ninter)}\big)$}
\end{algorithm}

In \cref{alg:girf} the $F$, $G$ and $P$ superscripts indicate filtered, guide and proposal particles, respectively.
The goal for the pseudocode in \cref{alg:girf}, and subsequent algorithms in this paper, is a succinct description of the logic of the procedure rather than a complete recipe for efficient coding.
Therefore, the pseudocode does not focus on opportunities for memory overwriting and vectorization, though these may be implemented in \pkg{\spatPomp} code.

We call the guide in \cref{alg:girf} a {\it bootstrap guide function} since it is based on resampling the Monte Carlo residuals calculated in step~\ref{alg:girf:guide:resid}.
Another option of a guide function in \code{girf} is the simulated moment guide function developed by \citet{park20} which uses the \code{eunit\_measure}, \code{vunit\_measure} and \code{munit\_measure} model components together with simulations to calculate the guide.
The expectation of Monte Carlo likelihood estimates does not depend on the guide function, so an inexact guide approximation may lead to loss of numerical efficiency but does not affect the consistency of the procedure.

The intermediate resampling is represented in \cref{alg:girf} by the loop of $\ninter=1,\dots,\Ninter$ in step~\ref{alg:girf:loop:ninter}.
The intermediate times are defined by $t_{\time,\ninter}=t_\time +
(t_{\time+1}-t_\time)\, \cdot \ninter\big/\Ninter$ and we write $\vec{X}_{\time,\ninter}=\vec{X}(t_{\time,\ninter})$.
The resampling weights (step~\ref{alg:girf:weights}) are defined in terms of guide function evaluations $\guideFunc^{P,\np}_{\time,\ninter}$.
The only requirement for the guide function to achieve unbiased estimates is that it satisfies
$\guideFunc^{F,\np}_{0,0}=1$
and
$\guideFunc^{P,j}_{\Time-1,\Ninter} =
  f_{\vec{Y}_{\Time}|\vec{X}_{\Time}} \big(
    \data{\vec{y}}_{\Time}\given \vec{X}^{F,\np}_{\Time-1,\Ninter}\giventh\vectheta
    \big)
$,
which is the case in \cref{alg:girf}.
The particular guide function calculated in step~\ref{alg:girf:guideFunc} evaluates particles using a prediction centered on a function
\[
\vec{\mu} (\vec{x},s,t\giventh\vectheta)\approx \E[\vec{X}(t)\given \vec{X}(s)=\vec{x}\giventh\vectheta].
\]
We call $\vec{\mu} (\vec{x},s,t\giventh\vectheta)$ a {\it deterministic trajectory} associated with $\vec{X}(t)$.
For a continuous time SpatPOMP model, this trajectory is typically the solution to a system of differential equations that define a vector field called the {\it skeleton} \citep{Tong1990}.
The skeleton is specified by a Csnippet filling the \code{skeleton} argument to \code{spatPomp()}.
The forecast spread around this deterministic prediction is given by the simulated bootstrap residuals constructed in step~\ref{alg:girf:guide:resid}.

\subsection[EnKF]{The ensemble Kalman filter (EnKF)}
\label{sec:enkf}

\begin{algorithm}[tb]
  \caption{\code{enkf(P,Np{\argequals}$J$)}, using notation from \cref{tab:notation} where \code{P} is a `\code{\spatPomp}' object equipped with
    \code{rprocess},
    \code{eunit\_measure},
    \code{vunit\_measure},
    \code{rinit},
    \code{coef},
    \code{obs}.
    \label{alg:enkf}
  }
  \KwIn{
    simulator for $f_{\vec{X}_{\time}|\vec{X}_{\time-1}}(\vec{x}_{\time} \given \vec{x}_{\time-1}\giventh\vectheta)$ and $f_{\vec{X}_0}(\vec{x}_0\giventh\vectheta)$;
    evaluator for $\emeasure_{\unit}(X_{\unit, \time},\vectheta)$ and $\vmeasure_{\unit}(X_{\unit,\time},\vectheta)$;
    parameter, $\vectheta$;
    data, $\data{\vec{y}}_{1:\Time}$;
    number of particles, $J$.
  }
  initialize filter particles,
  $\vec{X}_{0}^{F,\np}\sim {f}_{\vec{X}_{0}}\left(\mydot\giventh{\vectheta}\right)$
  for $\np$ in $\seq{1}{\Np}$
  \;
      \For{$\time\ \mathrm{in} \ \seq{1}{\Time}$}{
        prediction ensemble,
    $\vec{X}_{\time}^{P,\np}\sim {f}_{\vec{X}_{\time}|\vec{X}_{\time-1}}\big(\mydot|\vec{X}_{\time-1}^{F,\np};\vectheta\big)$
    for $\np$ in $\seq{1}{\Np}$
    \nllabel{alg:enkf:prediction}
    \;
        centered prediction ensemble, $\tilde{\vec{X}}_{\time}^{P,\np} =
        \vec{X}_{\time}^{P,\np} - \frac{1}{\Np}\sum_{\altNp=1}^{\Np}\vec{X}_{\time}^{P,\altNp}$
       for $\np$ in $\seq{1}{\Np}$
    \;
        forecast ensemble, $\vec{\hat{Y}}^{\np}_{\!\time}=\emeasure_{\unit}(X_{\unit,\time}^{P,\np},\vectheta)$
       for $\np$ in $\seq{1}{\Np}$
        \nllabel{alg:enkf:forecast}
        \;
	forecast mean, $\overline{\vec{Y}}_{\!\time}=\frac{1}{\Np}\sum_{\np=1}^{\Np}\vec{\hat{Y}}^{\np}_{\!\time}$
	\;
        centered forecast ensemble, $\vec{\tilde{Y}}^{\np}_{\time} =
        \vec{\hat{Y}}^{\np}_{\!\time} - \overline{\vec{Y}}_{\!\time}$
       for $\np$ in $\seq{1}{\Np}$
        \;
        forecast measurement variance,
	$R_{\unit,\altUnit} = \mathbbm{1}_{\unit,\altUnit} \,
	  \frac{1}{\Np}\sum_{\np=1}^{\Np}
	    \vmeasure_{\unit}\big(
	      \vec{X}_{\unit,\time}^{P,\np},\vectheta\big)$
         \nllabel{alg:enkf:cond:var}
	 for $\unit, \altUnit$ in $\seq{1}{\Unit}$
        \;
        forecast estimated covariance, $\Sigma_{Y}= \frac{1}{\Np-1}\sum_{\np=1}^{\Np}(\vec{\tilde{Y}}^{\np}_{\!\time})(\vec{\tilde{Y}}^{\np}_{\!\time})^T + R$
        \;
        prediction and forecast sample covariance, $\Sigma_{XY}=\frac{1}{\Np-1}\sum_{\np=1}^{\Np}(\tilde{\vec{X}}_{\time}^{P,\np})(\vec{\tilde{Y}}^{\np}_{\!\time})^T$
        \;
        Kalman gain, $K = \Sigma_{XY}\Sigma_{Y}^{-1}$
        \;
        artificial measurement noise, $\vec{\artNoise}_{\time}^{\np}\sim \normal(\vec{0},R)$
	for $\np$ in $\seq{1}{\Np}$
        \nllabel{alg:enkf:artificial:noise}
        \;
        errors, $\vec{r}_{\time}^{\np}= \vec{\hat{Y}}^{\np}_{\!\time} - \data{\vec{y}}_{\time}$
	for $\np$ in $\seq{1}{\Np}$
        \;
        filter update,
        $\vec{X}_{\time}^{F,\np} = \vec{X}_{\time}^{P,\np} +
              K\big( \vec{r}_{\time}^{\np}+\vec{\artNoise}_{\time}^{\np}\big)$
	for $\np$ in $\seq{1}{\Np}$
        \nllabel{alg:enkf:update}
	\;
	$\loglik_{\time}=\log \big[ \phi\big(\data{\vec{y}}_{\time} \giventh \overline{\vec{Y}}_{\!\time} , \Sigma_{Y} \big) \big]$ where $\phi(\cdot\giventh \vec{\mu},\Sigma)$ is the $\normal(\vec{\mu},\Sigma)$ density.
      }
  \KwOut{
    filter sample, $\vec{X}^{F,1:\Np}_{\time}$, for $n$ in $\seq{1}{N}$;
    log-likelihood estimate, $\loglik^{\mbox{\tiny{EnKF}}}=\sum_{\time=1}^{\Time} \loglik_{\time}$
  }
  \KwCplx{$\bigO(\Np\Unit\Time)$}
\end{algorithm}

Ensemble Kalman filter (EnKF) algorithms use observations to update simulations from the latent Markov model via an update rule based on a Gaussian conditional density \citep{evensen1994sequential,evensen96}.
The prediction step advances the Monte Carlo ensemble to the next observation time by using simulations from the postulated model
In the filtering step, the sample estimate of the state covariance matrix and the measurement variance are combined to update each ensemble member, using a rule that approximates the conditional distribution were the variables jointly Gaussian.

The \pkg{spatPomp} implementation of EnKF is described in \cref{alg:enkf}.
In step~\ref{alg:enkf:cond:var},  the conditional variance of the measurement at the current time step is approximated by constructing a diagonal covariance matrix whose diagonal elements are the sample average of the theoretical unit measurement variances at each unit.
This is written using an indicator function $\mathbbm{1}_{\unit,\altUnit}$ which takes value 1 if $\unit=\altUnit$ and 0 otherwise.
The \code{vunit\_measure} model component participates in this step whereas \code{eunit\_measure} specifies how we can construct forecast data (step \ref{alg:enkf:forecast}) that can be used to later update our prediction particles in step \ref{alg:enkf:update}.
In step \ref{alg:enkf:artificial:noise} we add artificial measurement error to arrive at a consistent sample covariance for the filtering step \citep{evensen1994sequential,evensen96}, writing $\normal(\vec{\mu},\Sigma)$ for independent draws from a multivariate normal random variable with mean $\vec{\mu}$ and variance matrix $\Sigma$.

EnKF achieves good dimensional scaling relative to PF by replacing the resampling step with an update rule inspired by a Gaussian approximation.
If the number of units exceeds the number of particles in the ensemble, regularization is required when estimating the covariance matrix.
Currently, \pkg{\spatPomp} has focused on systems for which this issue does not arise.

Our EnKF implementation supposes we have access to the measurement mean function, $\emeasure_{\unit,\time}(x,\vectheta)$, and the measurement variance, $\vmeasure_{\unit}(x,\vectheta)$, defined in \Cref{tab:notation}.
For common choices of measurement model, such as Gaussian or negative binomial, $\emeasure_{\unit,\time}$ and $\vmeasure_{\unit,\time}$ are readily available.
\Cref{sec:measles} provides an example of simple Csnippets for \code{eunit\_measure} and \code{vunit\_measure}.
In general, the functional forms of $\emeasure_{\unit,\time}$ and $\vmeasure_{\unit,\time}$ may depend on $\unit$ and $\time$, or on covariate time series. 

\subsection[BPF]{Block particle filter}
\label{sec:bpfilter}

\begin{algorithm}[t]
 \caption{\code{bpfilter(P,Np{\argequals}$\Np\!$,block\_list{\argequals}$\blocklist$)}
    using notation from \cref{tab:notation} where \code{P} is a
     `\code{\spatPomp}' object equipped with
     \code{rprocess},
     \code{dunit\_measure},
     \code{rinit},
     \code{obs},
     \code{coef}.
   }\label{alg:bpfilter}
  \KwIn{
    simulator for $f_{\vec{X}_{\time}|\vec{X}_{\time-1}}(\vec{x}_{\time}\given \vec{x}_{\time-1}\giventh\vectheta)$ and $f_{\vec{X}_0}(\vec{x}_{0}\giventh\vectheta)$;
   number of particles, $\Np$;
    evaluator for $f_{{Y}_{\unit,\time}|{X}_{\unit,\time}}({y}_{\unit,\time}\given {x}_{\unit,\time}\giventh\vectheta)$;
    data, $\data{\vec{y}}_{1:\Time}$;
    parameter, $\vectheta$;
     blocks, $\blocklist_{1:\Block}$;
  }
initialization, $\vec{X}_{0}^{F,\np}\sim {f}_{\vec{X}_{0}}\left(\mydot\giventh{\vectheta}\right)$
for $\np$ in $\seq{1}{\Np}$
\;
\For{$\time\ \mathrm{in}\ \seq{1}{\Time}$}{ % For each time step
  prediction,
  $\vec{X}_{\time}^{P,\np}\sim {f}_{\vec{X}_{\time}|\vec{X}_{\time-1}}\big(\mydot|\vec{X}_{\time-1}^{F,\np};{\vectheta}\big)$
  for $\np$ in $\seq{1}{\Np}$
    \;
    block weights,
          $\displaystyle \blockweight_{\block,\time}^{\np}=
            \prod_{\unit \in \blocklist_{\block}}
            f_{Y_{\unit,\time}|X_{\unit,\time}}
            \big(
              \data{y}_{\unit,\time}\given X^{P,\np}_{\unit,\time} \giventh \vectheta
            \big)$
    for $\np$ in $\seq{1}{\Np}$, $\block$ in $\seq{1}{\Block}$
          \nllabel{alg:bpfilter:blockweights}\;
   resampling indices,
      $\resampleIndex^{\np}_{\block,\time}$ with
      $\prob\left[\resampleIndex^{\np}_{\block,\time}=i \right] =
      {\blockweight}^{i}_{\block,\time}
      \Big/\sum_{q=1}^{\Np}\blockweight^{q}_{\block,\time}$
      for $\np$ in $\seq{1}{\Np}$, $\block$ in $\seq{1}{\Block}$
    \;
    resample,
      $X_{\blocklist_{\block},\time}^{F,\np}=
      X_{\blocklist_{\block},\time}^{P,\resampleIndex_{\np,\block}}$
      for $\np$ in $\seq{1}{\Np}$, $\block$ in $\seq{1}{\Block}$
      \nllabel{alg:bpfilter:resample}
    \;
}%End timestep for Loop
\KwOut{
    log-likelihood,
      $\loglik^{\mbox{\tiny{BPF}}}(\vectheta)= \sum_{\time=1}^{\Time}
        \sum_{\block=1}^{\Block}
\log\Big(
        \frac{1}{\Np}
        \sum_{\np=1}^{\Np}
          \blockweight^{\np}_{\block,\time}
      \Big)$, filter particles $\vec{X}^{F,1:\Np}_{1:\Time}$
}
\KwCplx{$\bigO(\Np \Unit \Time)$}
\end{algorithm}

\Cref{alg:bpfilter} is an implementation of the block particle filter \citep[BPF][]{rebeschini15}, also called the factored particle filter \citep{ng02}.
BPF partitions the units into a collection of blocks, $\blocklist_1,\dots,\blocklist_{\Block}$, such that each unit is a member of exactly one block.
BPF generates proposal particles by simulating from the joint latent process across all blocks, exactly as the particle filter does.
However, the resampling in the filtering step is carried out independently for each block, using weights corresponding only to the measurements in the block.
Different proposal particles may be successful for different blocks, and the block resampling allows the filter particles to paste together these successful proposals.
This avoids the curse of dimensionality, while introducing an approximation error that may be large or small depending on the model under consideration.

The user has a choice of specifying the blocks using either the \code{block\_list} argument or \code{block\_size}, but not both.
\code{block\_list} takes a \class{list} object where each entry is a vector representing the units in a block.
\code{block\_size} takes an integer and evenly partitions $1\mycolon\Unit$ into blocks of size approximately \code{block\_size}.
For example, if there are 4 units, executing \code{bpfilter} with \code{block\_size=2} is equivalent to setting \code{block\_list=list(c(1,2),c(3,4))}.

\subsection[\ABF]{Adapted bagged filter (\ABF)}
\label{sec:abf}

The adapted bagged filter \citep{ionides23} combines many independent particle filters.
This is called \emph{bagging}, (\emph{b}ootstrap \emph{agg}regat\emph{ing}), since a basic particle filter is also called a bootstrap filter.
The adapted distribution is the conditional distribution of the latent process given its current value and the subsequent observation \citep{johansen08}.
In the adapted bagged filter, each bootstrap replicate makes a Monte Carlo approximation to a draw from the adapted distribution.
Thus, in the pseudocode of \cref{alg:abf}, $\vec{X}^{\IF,\rep}_{0:\Time}$ is a Monte Carlo sample targeting the adapted sampling distribution,
\begin{equation}
  \label{adapteddistribution}
f_{\vec{X}_0}(\vec{x}_0\giventh\vectheta)\prod_{\time=1}^\Time
f_{\vec{X}_{\time}|\vec{Y}_{\time},\vec{X}_{\time-1}}
(\vec{x}_{\time}\given \data{\vec{y}}_{\time},\vec{x}_{\time-1} \param\vectheta).
\end{equation}
Each adapted simulation replicate is constructed by importance sampling using proposal particles $\{\vec{X}^{\IP,\rep,\np}_{\time}\}$.
The ensemble of adapted simulation replicates are then weighted using data in a spatiotemporal neighborhood of each observation to obtain a locally combined Monte Carlo sample targeting the filter distribution, with some approximation error due to the finite spatiotemporal neighborhood used.
This local aggregation of the bootstrap replicates also provides an evaluation of the likelihood function.

On a given bootstrap replicate $\rep$ at a given time $\time$, all the adapted proposal particles $\vec{X}^{\IP,\rep,1:\Np}_{\time}$ in step~\ref{alg:abf:adapted:proposals} are necessarily close to each other in state space because they share the parent particle $\vec{X}^{\IF,\rep}_{\time-1}$.
This reduces imbalance in the adapted weights in step~\ref{alg:abf:adapted:weights}, which helps to battle the curse of dimensionality that afflicts importance sampling.
The combination of the replicates for the filter estimate in step~\ref{alg:abf:local:filter} is carried out using only weights in a spatiotemporal neighborhood, thus avoiding the curse of dimensionality.
For any point $(\unit, \time)$, the neighborhood $B_{\unit,\time}$ should be specified as a subset of $A_{\unit,\time}=\{(\tilde\unit,\tilde\time): \mbox{$\tilde\time < \time$ or ($\tilde\unit< \unit$ and $\tilde\time = \time$)} \}$.
If the model has a mixing property, meaning that conditioning on the observations in the neighborhood $B_{\unit,\time}$ is negligibly different from conditioning on the full set $A_{\unit,\time}$, then the approximation involved in this localization is adequate.

\begin{algorithm}[t]
  \caption{\code{abf(P,replicates{\argequals}$\Rep$,Np{\argequals}$J$,nbhd=$B_{\unit,\time}$)},
     using notation from \cref{tab:notation} where \code{P} is a
     `\code{\spatPomp}' object equipped with
     \code{rprocess},
     \code{dunit\_measure},
    \code{rinit},
    \code{obs},
    \code{coef}.
  }\label{alg:abf}
  \KwIn{
    simulator for $f_{\vec{X}_{\time}|\vec{X}_{\time-1}}(\vec{x}_{\time}\given \vec{x}_{\time-1}\giventh\vectheta)$ and $f_{\vec{X}_0}(\vec{x}_0\giventh\vectheta)$;
    evaluator for $f_{{Y}_{\unit,\time}|{X}_{\unit,\time}}({y}_{\unit,\time}\given {x}_{\unit,\time}\giventh\vectheta)$;
    data, $\data{\vec{y}}_{1:\Time}$;
    parameter, $\vectheta$;
    number of particles per replicate, $\Np$;
    number of replicates, $\Rep$;
    neighborhood structure, $B_{\unit,\time}$
  }
initialize adapted simulation, $\vec{X}^{\IF,\rep}_{0} \sim f_{\vec{X}_0}(\cdot\giventh\vectheta)$
  for $\rep$ in $\seq{1}{\Rep}$ \;
        \nllabel{alg:abf:for:n}
\For{$\time\ \mathrm{in}\ \seq{1}{\Time}$}{
  proposals,
    $\vec{X}_{\time}^{\IP,\rep,\np} \sim
    f_{\vec{X}_{\time}|\vec{X}_{\time-1}}
    \big( \mydot \given \vec{X}^{\IF,\rep}_{\time-1} \giventh \vectheta \big)$
  for $\rep$ in $\seq{1}{\Rep}$, $\np$ in $\seq{1}{\Np}$
  \nllabel{alg:abf:adapted:proposals}\;
%  unit measurement density:
  $\unitWeight_{\unit,\time}^{\rep,\np} =
    f_{Y_{\unit,\time}|X_{\unit,\time}}
    \big (\data{y}_{\unit,\time}\given X^{\IP,\rep,\np}_{\unit,\time} \giventh \vectheta\big)$
    for $\unit$ in $\seq{1}{\Unit}$, $\rep$ in $\seq{1}{\Rep}$, $\np$ in $\seq{1}{\Np}$
  \;
  adapted resampling weights,
  $w^{\IF,\rep,\np}_{\time} =
    \prod_{\unit=1}^{\Unit} \unitWeight^{\rep,\np}_{\unit,\time}$
    for $\unit$ in $\seq{1}{\Unit}$, $\rep$ in $\seq{1}{\Rep}$,
    $\np$ in $\seq{1}{\Np}$
  \nllabel{alg:abf:adapted:weights}\;
  set $\vec{X}^{\IF,\rep}_{\time} = \vec{X}^{\IP,\rep,\np}_{\time}$
    with probability $w^{\IF,\rep,\np}_{\time}
  \left(
  \sum_{\altNp=1}^{\Np} w^{\IF,\rep,\altNp}_{\time}
  \right)^{-1}$
  for $\rep$ in $\seq{1}{\Rep}$
  \;
%local prediction weights,
  $w^{\LCP,\rep,\np}_{\unit,\time}= \displaystyle
  \prod_{\altTime=1}^{\time-1}
  \left[
    \frac{1}{\Np}\sum_{\altNp=1}^{\Np}
    \hspace{2mm}
       \prod_{(\altUnit,\altTime)\in B_{\unit,\time}}
         \hspace{-2mm}
        \unitWeight_{\altUnit,\altTime}^{\rep,\altNp}
  \right] \prod_{(\altUnit,\time)\in B_{\unit,\time}}
         \hspace{-2mm}
         \unitWeight_{\altUnit,\time}^{\rep,\np}$
      for $\unit$ in $\seq{1}{\Unit}$, $\rep$ in $\seq{1}{\Rep}$, $\np$ in $\seq{1}{\Np}$
  \nllabel{alg:abf:lp:weights} \nllabel{alg:abf:end:n}
}
filter weights,
$\displaystyle w^{\LCF,\rep,\np}_{\unit,\time}=
 \frac{
   \unitWeight_{\unit,\time}^{\rep,\np} \, \, w^{\LCP,\rep,\np}_{\unit,\time}
 }{
    \sum_{p=1}^{\Rep}
    \sum_{\altNp=1}^{\Np}
    w^{\LCP,p,\altNp}_{\unit,\time}
}
  $
  for $\unit$ in $\seq{1}{\Unit}$, $\time$ in $\seq{1}{\Time}$, $\rep$ in $\seq{1}{\Rep}$, $\np$ in $\seq{1}{\Np}$
  \nllabel{alg:abf:lf:weights} \;
conditional log-likelihood,
${\loglik}_{\unit,\time}=
  \log\left(
    \sum_{\rep=1}^{\Rep}\sum_{\np=1}^{\Np} w^{\LCF,\rep,\np}_{\unit,\time}
  \right)$
  for $\unit$ in $\seq{1}{\Unit}$, $\time$ in $\seq{1}{\Time}$\;
% Local filter:
set $X^{\LCF,\np}_{\unit,\time} = {X}^{\IP,\rep,k}_{\unit,\time}$ with probability  $w^{\LCF,\rep,k}_{\unit,\time} \, e^{-{\loglik}_{\unit,\time}}$
    for $\unit$ in $\seq{1}{\Unit}$, $\time$ in $\seq{1}{\Time}$, $\np$ in $\seq{1}{\Np}$
  \nllabel{alg:abf:local:filter}
  \;
\KwOut{
filter particles, $\vec{X}^{\LCF,1:\Np}_{\time}$, for $\time$ in $\seq{1}{\Time}$;
log-likelihood, $\loglik^{\mbox{\tiny{ABF}}}= \sum_{\time=1}^\Time\sum_{\unit=1}^\Unit \loglik_{\unit,\time}$
}
  \KwCplx{$\bigO(\Rep \Np \Unit \Time)$}
\end{algorithm}

Steps~\ref{alg:abf:for:n} through~\ref{alg:abf:end:n} do not involve interaction between replicates and therefore iteration over $\rep$ can be carried out in parallel.
If a parallel back-end has been set up by the user, the \code{abf} method will parallelize computations over the replicates using multiple cores.
The user can register a parallel back-end using the \code{doParallel} package \citep{R:foreach,R:doParallel} prior to calling \code{abf}.
\begin{knitrout}
\definecolor{shadecolor}{rgb}{1, 1, 1}\color{fgcolor}\begin{kframe}
\begin{verbatim}
R> library("doParallel")
R> registerDoParallel(detectCores()) 
\end{verbatim}
\end{kframe}
\end{knitrout}

The neighborhood is supplied via the \code{nbhd} argument to \code{abf} as a function which takes a point in space-time, $(\unit, \time)$, and returns a list of points in space-time which correspond to $B_{\unit,\time}$.
An example with $B_{\unit,\time} = \{(u-1,n),(u,n-1)\}$ follows.
\begin{knitrout}
\definecolor{shadecolor}{rgb}{1, 1, 1}\color{fgcolor}\begin{kframe}
\begin{verbatim}
R> example_nbhd <- function(object, unit, time){
+    nbhd_list = list()
+    if(time>1) nbhd_list <- c(nbhd_list, list(c(unit, time-1)))
+    if(unit>1) nbhd_list <- c(nbhd_list, list(c(unit-1, time)))
+    return(nbhd_list)
+  }
\end{verbatim}
\end{kframe}
\end{knitrout}

ABF can be combined with the guided intermediate resampling technique used by GIRF to give an algorithm called ABF-IR \citep{ionides23} implemented as \code{abfir}.

\subsection{Considerations for choosing a filter}

Of the four filters described above, only GIRF provides an unbiased estimate of the likelihood.
However, GIRF has a relatively weak theoretical scaling support, beating the curse of dimensionality only in the impractical situation of an ideal guide function \citep{park20}.
EnKF, ABF and BPF gain scalability by making different approximations that may or may not be appropriate for a given situation.
The choice of filter in a particular application is primarily an empirical question, and \pkg{spatPomp} facilitates a responsible approach of trying multiple options.
Nevertheless, we offer some broad guidance.
EnKF has low variance but is relatively sensitive to deviations from normality; in examples where this is not a concern, such as the example in ~\cref{sec:brownian}, EnKF can be expected to perform well.
BPF can break conservation laws satisfied by the latent process, such as a constraint on the total population in all units;
ABF satisfies such constraints but has been found to have higher variance than BPF on some benchmark problems \citep{ionides23}.
For the measles model built by \code{measles()}, BPF and ABF have been found to perform better than EnKF and GIRF \citep{ionides23}.
For the Lorenz-96 example built by \code{lorenz()}, GIRF and BPF perform well \citep{ionides23}.

\section{Inference for SpatPOMP models}
\label{sec:inference}

We focus on iterated filtering methods \citep{ionides15} which provide a relatively simple way to coerce filtering algorithms to carry out parameter inference, applicable to the general class of SpatPOMP models considered by \pkg{spatPomp}.
The main idea of iterated filtering is to extend a POMP model to include dynamic parameter perturbations.
Repeated filtering, with parameter perturbations of decreasing magnitude, approaches the maximum likelihood estimate.
Here, we present iterated versions of GIRF, EnKF, BPF, and the unadapted bagged filter (UBF), a version of ABF with $\Np=1$. These algorithms are known as {\iGIRF} \citep{park20}, {\iEnKF} \citep{li20}, IBPF \citep{ning23-ibpf,ionides22} and {\iUBF} \citep{ionides23} respectively.
SpatPOMP model estimation is an active area for research \citep[for example,][]{katzfuss20} and \pkg{spatPomp} provides a platform for developing new statistical methods and testing them on a range of models.

\begin{algorithm}[p]
  \caption{\code{igirf(P,}\code{params{\argequals}$\vectheta_0$,}\code{Ngirf{\argequals}$\Nit$,}\code{Np{\argequals}$\Np$,}\code{Ninter{\argequals}$\Ninter$,}\code{Nguide{\argequals}$\Nguide$,}\code{Lookahead{\argequals}$\Lookahead$,}
  \code{rw.sd{\argequals}$\sigma_{0:\Time,1:\Thetadim}$,}\code{cooling.fraction.50{\argequals}$a$)}
     using notation from \cref{tab:notation} where \code{P} is a `\code{\spatPomp}' object equipped with
     \code{rprocess},
     \code{dunit\_measure},
     \code{skeleton},
     \code{rinit},
     \code{obs}.
  }\label{alg:igirf}
  \KwIn{
    simulator for $f_{\vec{X}_{\time}|\vec{X}_{\time-1}}(\vec{x}_{\time} \given \vec{x}_{\time-1}\giventh\vectheta)$ and $f_{\vec{X}_0}(\vec{x}_0\giventh\vectheta)$;
    evaluator for $f_{{Y}_{\unit,\time}|{X}_{\unit,\time}}({y}_{\unit,\time}\given {x}_{\unit,\time}\giventh\vectheta)$,
and $\vec{\mu}(\vec{x},s,t\giventh\vectheta)$;
    data, $\data{\vec{y}}_{1:\Time}$;
    starting parameter, $\vectheta_0$;
    iterations, $\Nit$;
    particles, $J$;
    lookahead lags, $\Lookahead$;
    intermediate timesteps, $\Ninter$;
    random walk intensities, $\sigma_{0:\Time,1:\Thetadim}$;
    cooling fraction in 50 iterations, $a$.
    }
  \KwIndices{free indices are implicit `for' loops, calculated for $\np\ \text{in}\ \seq{1}{\Np}$, $\nguide\ \text{in}\ \seq{1}{\Nguide}$,  $\lookahead\ \text{in}\ \seq{(\time+1)}{\lookaheadEnd}$, $\unit\ \text{in}\ \seq{1}{\Unit}$, $\thetadim, \thetadim^\prime \ \text{in}\ \seq{1}{\Thetadim}$.}
  \BlankLine
  initialize parameters,
    $\vecTheta^{F,0,\np}_{\Time-1,\Ninter} =  \theta_0$
  \;
\For{$\nit\ \mathrm{in}\ \seq{1}{\Nit}$ \nllabel{alg:igirf:loop:m} }{
  initialize parameters,
    $\vecTheta^{F,\nit,\np}_{0,0} \sim \normal \big(\vecTheta^{F,\nit-1,\np}_{\Time-1,\Ninter} \, ,
	  a^{2\nit/50} \, \Sigma_{\text{ivp}}
	  \big)$ for $\big[\Sigma_{\text{ivp}}\big]_{\thetadim,\thetadim^\prime} = \sigma_{\text{ivp},\thetadim}^2 \mathbbm{1}_{\thetadim=\thetadim^\prime}$
  \nllabel{alg:igirf:initialize}
  \;
  initialize filter particles,
  simulate $\vec{X}_{0,0}^{F,\np}\sim {f}_{\vec{X}_{0}}\left(\mydot\giventh{\vecTheta^{F,\nit,\np}_{0,0}}\right)$ and set $\guideFunc^{F,\np}_{\time,0}=1$
  \;
  \For{$\time\ \mathrm{in}\ \seq{0}{\Time-1}$}{
    guide simulations,
      $\vec{X}_{\lookahead}^{G,\np,\nguide}
      \sim
      {f}_{\vec{X}_{\lookahead}|\vec{X}_{\time}}
      \big( \mydot|\vec{X}_{\time,0}^{F,\np} \giventh {\vecTheta_{\time,0}^{F,\nit,\np}} \big)$
      \nllabel{alg:igirf:guide:sim}\;
    guide residuals, $\vec{\epsilon}^{\np,\nguide}_{0,\lookahead}=
        \vec{X}_{\lookahead}^{G,\np,\nguide} -
	\vec{\mu}\big(
	  \vec{X}^{F,\np}_{\time,0},t_{\time},t_{\lookahead}
	  \giventh \vecTheta^{F,\nit,\np}_{\time,0}
	\big)$
        \nllabel{alg:igirf:guide:resid}\;
    \nllabel{alg:igirf:loop:ninter}
    \For{$\ninter  \ \mathrm{in}\ \seq{1}{\Ninter}$}{
      perturb parameters,
        $\vecTheta^{P,\nit,\np}_{\time,\ninter}\sim \normal \big(
          \vecTheta^{F,\nit,\np}_{\time,\ninter-1} \, ,
	  a^{2\nit/50} \, \Sigma_{\time}
	  \big)$ for $\big[\Sigma_{\time}\big]_{\thetadim,\thetadim^\prime} = \sigma_{\time,\thetadim}^2 \mathbbm{1}_{\thetadim=\thetadim^\prime}/\Ninter$
  \;
      prediction simulations,
        ${\vec{X}}_{\time,\ninter}^{P,\np}
          \sim {f}_{{\vec{X}}_{\time,\ninter}|{\vec{X}}_{\time,\ninter-1}}
          \big(\mydot|{\vec{X}}_{\time,\ninter-1}^{F,\np} \giventh {\vecTheta_{\time,\ninter}^{P,\nit,\np}}\big)$
          \nllabel{alg:igirf:predictions}\;
      deterministic trajectory,
        $\vec{\mu}^{P,\np}_{\time,\ninter,\lookahead}
           = \vec{\mu} \big( \vec{X}^{P,\np}_{\time,\ninter},t_{\time,\ninter},t_{\lookahead} \giventh \vecTheta_{\time,\ninter}^{P,\nit,\np} \big)$
        \nllabel{alg:igirf:guide:skeleton}
	\;
      pseudo guide simulations,
        $\hat{\vec{X}}^{\np,\nguide}_{\time,\ninter,\lookahead} =
        \vec{\mu}^{P,\np}_{\time,\ninter,\lookahead} +
        \vec{\epsilon}^{\np,\nguide}_{\ninter-1,\lookahead} -
        \vec{\epsilon}^{\np,\nguide}_{\ninter-1,\time+1} +
        {\textstyle \sqrt{
            \frac{t_{\time+1}-t_{\time,\ninter}}{t_{\time+1}-t_{\time,0}}}
        }	\,
        \vec{\epsilon}^{\np,\nguide}_{\ninter-1,\time+1}$
        \;
      discount factor,
        $\eta_{\time, \ninter,\lookahead}
          = 1-(t_{\time+\lookahead}-t_{\time,\ninter})/\{(t_{\time+\lookahead}-t_{\max(\time+\lookahead-\Lookahead,0)})\cdot(1+\mathbbm{1}_{\Lookahead=1})\}$\;
% Guide function:
        $ \displaystyle
\guideFunc^{P,\np}_{\time,\ninter}=
          \prod_{\lookahead=\time+1}^{\lookaheadEnd}
          \prod_{\unit=1}^{\Unit}
          \left[
	  \frac{1}{\Nguide}
          \sum_{\nguide=1}^{\Nguide}
          f_{Y_{\unit,\lookahead}|X_{\unit,\lookahead}}
          \Big(
            \data{y}_{\unit,\lookahead}\given \hat{X}^{\np,\nguide}_{\unit,\time,\ninter,\lookahead}
	     \giventh \vecTheta_{\time,\ninter}^{P,\nit,\np}
          \Big)
          \right]^{\eta_{\time, \ninter,\lookahead}}$
        \nllabel{alg:igirf:guideFunc}\;
        $w^{\np}_{\time,\ninter}=\left\{
          \begin{array}{ll}
          f_{\vec{Y}_{\time}|\vec{X}_{\time}} \big(
            \vec{y}_{\time}\given \vec{X}^{F,\np}_{\time,\ninter-1}\giventh\vecTheta_{\time,\ninter-1}^{F,\nit,\np}
            \big) \,\,
         \guideFunc^{P,\np}_{\time,\ninter}
         \Big/ \guideFunc^{F,\np}_{\time,\ninter-1}
          & \mbox{if $s=1$, $n\neq 0$} \\
            \guideFunc^{P,\np}_{\time,\ninter}
             \Big/ \guideFunc^{F,\np}_{\time,\ninter-1}
          & \mbox{else}
          \end{array} \right.$
        \nllabel{alg:igirf:weights}\;
      normalized weights,
        $\tilde{w}^{\np}_{\time,\ninter}= w^{\np}_{\time,\ninter}\Big/\sum_{\altNp=1}^{\Np}w^{\altNp}_{\time,\ninter}$\;
      resampling indices,
        $\resampleIndex_{1:\Np}$
        with
        $\prob\left[\resampleIndex_{\np}=\altNp\right] =\tilde{w}^{\altNp}_{\time,\ninter}$
        \nllabel{alg:igirf:systematic}\;
      set
        $\vec{X}_{\time,\ninter}^{F,\np}=\vec{X}_{\time,\ninter}^{P,\resampleIndex_{\np}}\,$,
        $\; \guideFunc^{F,\np}_{\time,\ninter}= \guideFunc^{P,\resampleIndex_{\np}}_{\time,\ninter}\,$,
	        $\; \vec{\epsilon}^{\np,\nguide}_{\ninter,\lookahead}=
	\vec{\epsilon}^{\resampleIndex_{\np},\nguide}_{\ninter-1,\lookahead}$,
        $\; \vecTheta_{\time,\ninter}^{F,\nit,\np} =
             \vecTheta_{\time,\ninter}^{P,\nit,\resampleIndex_{\np}}$
        \nllabel{alg:igirf:resample}
    }
  }
}
  \KwOut{
    Iterated GIRF parameter swarm, $\vecTheta_{\Time-1,S}^{F,\Nit,1:\Np}$ \;
    Monte Carlo maximum likelihood estimate: $\frac{1}{\Np} \sum_{\np=1}^{\Np}\vecTheta^{F,\Nit,\np}_{\Time-1,S}$
  }
  \KwCplx{$\bigO\big({\Nit\Np\Lookahead\Unit\Time(\Nguide+\Ninter)}\big)$}
\end{algorithm}

\subsection[Iterated GIRF]{Iterated GIRF for parameter estimation}
\label{sec:igirf}

\Cref{alg:igirf} describes \code{igirf()}, the \pkg{\spatPomp} implementation of IGIRF.
This algorithm carries out the IF2 algorithm of \citet{ionides15} with filtering carried out by GIRF, therefore its implementation combines the \code{mif2} function in \pkg{pomp} with \code{girf} (\cref{alg:girf}).
To unclutter the pseudocode, in this section we use the convention that a free subscript or superscript indicates an implicit `for' loop over all values in the index range.
Here, a numeric index is called ``free'' if its value is not explicitly specified by the code.
For example, in Line~\ref{alg:igirf:initialize} of \cref{alg:igirf} there is an implicit loop over all values of $\np$ in $\seq{1}{\Np}$, but not over $\nit$ since that was specified explicitly in Line~\ref{alg:igirf:loop:m}.

The quantity $\vecTheta^{P,\nit,\np}_{\time,\ninter}$ gives a perturbed parameter vector for $\vectheta$ corresponding to particle $\np$ on iteration $\nit$ at the $\ninter^{\text{th}}$ intermediate time between $\time$ and $\time+1$.
The perturbations in \cref{alg:igirf} are taken to follow a multivariate normal distribution, with a diagonal covariance matrix scaled by $\sigma_{\time,\thetadim}$.
Normally distributed perturbations are not theoretically required, but are a common choice in practice.
The \code{igirf} function permits perturbations to be carried out on a transformed scale, specified using the \code{partrans} argument, to accommodate situations where normally distributed perturbations are more natural on the log or logistic scale, or any other user-specified scale.
For regular parameters, i.e. parameters that are not related to the initial conditions of the dynamics, it may be appropriate to set the perturbation scale independent of $\time$.
If parameters are transformed so that a unit scale is relevant, for example using a logarithmic transform for non-negative parameters, an appropriate default value is $\sigma_{\time,\thetadim}=0.02$.
Initial value parameters (IVPs) are those that determine only the latent state at time $t_0$, and these should be perturbed only at the beginning of each iteration $\nit$.
The matrix $\sigma_{0:\Time,1:\Thetadim}$ can be constructed using the \code{rw\_sd} function, which simplifies the construction of the \code{rw.sd} argument for regular parameters and IVPs.
The \code{cooling.fraction.50} argument takes the fraction of \code{rw.sd} by which to perturb the parameters after 50 iterations of \code{igirf}.
If using the default geometric cooling schedule, a value of \code{cooling.fraction.50=0.5} means that the perturbation standard deviation decreases roughly 1\% per iteration.

\subsection[Iterated EnKF]{Iterated EnKF for parameter estimation}
\label{sec:ienkf}

\begin{algorithm}[tbp]
  \caption{\code{ienkf(P,}\code{params{\argequals}$\vectheta_0$,}\code{Nenkf{\argequals}$\Nit$,}\code{cooling.fraction.50{\argequals}$a$,}\code{rw.sd{\argequals}$\sigma_{0:\Time,1:\Thetadim}$,}
  \code{Np{\argequals}$J$)},
     using notation from \cref{tab:notation} where \code{P} is a `\code{\spatPomp}' object equipped with
     \code{rprocess},
     \code{eunit\_measure},
     \code{vunit\_measure},
     \code{rinit},
     and \code{obs}.
     \label{alg:ienkf}
  }
  \KwIn{
    simulator for $f_{\vec{X}_{\time}|\vec{X}_{\time-1}}(\vec{x}_{\time} \given \vec{x}_{\time-1}\giventh\vectheta)$ and $f_{\vec{X}_0}(\vec{x}_0\giventh\vectheta)$;
    evaluator for $\emeasure_{\unit,\time}(x,\vectheta)$ and $\vmeasure_{\unit,\time}(x,\vectheta)$;
    data, $\data{\vec{y}}_{1:\Time}$;
    number of particles, $J$;
    number of iterations, $\Nit$;
    starting parameter, $\vectheta_0$;
    random walk intensities, $\sigma_{0:\Time,1:\Thetadim}$;
    cooling fraction in 50 iterations, $a$.
  }
  \KwIndices{free indices are implicit `for' loops, calculated for $\np\ \text{in}\ \seq{1}{\Np}$, $\unit$ and $\altUnit$ in $\seq{1}{\Unit}$, $\thetadim$ and $\thetadim^\prime$ in $\seq{1}{\Thetadim}$.}
  \BlankLine
initialize parameters,
    $\vecTheta^{F,0,\np}_{\Time} =\vectheta_{0}$
\;
\For{$\nit\ \mathrm{in}\ \seq{1}{\Nit}$}{
  initialize parameters,
      $\Theta^{F,\nit,\np}_{0}\sim \normal
        \big(\Theta^{F,\nit-1,\np}_{\Time} \,\,  , \,
	  a^{2\nit/50} \, \Sigma_{0}
	\big)$ for $\big[\Sigma_{\time}\big]_{\thetadim,\thetadim^\prime} = \sigma_{\time,\thetadim}^2 \mathbbm{1}_{\thetadim=\thetadim^\prime}$
  \;
  initialize filter particles,
  simulate $\vec{X}_{0}^{F,\np}\sim {f}_{\vec{X}_{0}}\left(\mydot\giventh{\vecTheta^{F,\nit,\np}_{0}}\right)$.\;
      \For{$\time\ \mathrm{in} \ \seq{1}{\Time}$}{
        perturb parameters,
	  $\Theta_{\time}^{P,\nit,\np} \sim  \normal \big(
          \Theta^{F,\nit,\np}_{\time-1} \, \, , \,
	   a^{2\nit/50} \, \Sigma_{\time}
          \big)$
    \;
        prediction ensemble,
    $\vec{X}_{\time}^{P,\np}\sim {f}_{\vec{X}_{\time}|\vec{X}_{\time-1}}\big(\mydot|\vec{X}_{\time-1}^{F,\np};{\vecTheta_{\time}^{P,\nit,\np}}\big)$
    \nllabel{alg:ienkf:step1}
    \;
        process and parameter ensemble, $\vec{Z}_{\time}^{P,\np} =\colvec{\vec{X}^{P,\np}_{\time}}{\vecTheta^{P,\nit,\np}_{\time}}$
        \;
        centered process and parameter ensemble, $\tilde{\vec{Z}}_{\time}^{P,\np} =
        \vec{Z}_{\time}^{P,\np} - \frac{1}{\Np}\sum_{\altNp=1}^{\Np}\vec{Z}_{\time}^{P,\altNp}$
\;
        forecast ensemble, $\hat{\vec{Y}}^{\np}_{\time}$, defined by
	${\hat{Y}}_{\unit,\time}^{\np}=
	  \emeasure_{\unit}(X_{\unit,\time}^{P,\np},\vecTheta_{\time}^{P,\nit,\np})$
        \nllabel{alg:ienkf:forecast:ensemble}
        \;
        centered forecast ensemble, $\vec{\tilde{Y}}_{\time}^{\np}=
        \vec{\hat{Y}}_{\time}^{\np} - \frac{1}{\Np}\sum_{\altNp=1}^{\Np}\vec{\hat{Y}}_{\time}^{\altNp}$
        \;
        forecast measurement variance,
	$R_{\unit,\altUnit} = \mathbbm{1}_{\unit,\altUnit} \,
	  \frac{1}{\Np}\sum_{\np=1}^{\Np}
	    \vmeasure_{\unit}(
	      \vec{X}_{\unit,\time}^{P,\np},
	      \vecTheta_{\time}^{P,\nit,\np})$
        \;
        forecast sample covariance, $\Sigma_{Y}= \frac{1}{\Np-1}\sum_{\np=1}^{\Np}(\vec{\tilde{Y}}_{\time}^{\np})(\vec{\tilde{Y}}_{\time}^{\np})^T + R$
        \;
        prediction and forecast sample covariance, $\Sigma_{ZY}=\frac{1}{\Np-1}\sum_{\np=1}^{\Np}(\tilde{\vec{Z}}_{\time}^{P,\np})(\vec{\tilde{Y}}_{\time}^{\np})^T$
        \;
        Kalman gain: $K = \Sigma_{ZY}\Sigma_{Y}^{-1}$
        \;
        artificial measurement noise, $\vec{\artNoise}_{\time}^{\np}\sim \normal(\vec{0},R)$
        \;
        errors, $\vec{r}_{\time}^{\np}= \vec{\hat{Y}}_{\time}^{\np} - \data{\vec{y}}_{\time}$
	\nllabel{alg:ienkf:r}
        \;
        filter update:
        $\vec{Z}_{\time}^{F,\np} = \colvec{\vec{X}^{F,\np}_{\time}}{\vecTheta^{F,\nit,\np}_{\time}} = \vec{Z}_{\time}^{P,\np} +
              K\big( \vec{r}_{\time}^{\np}+\vec{\artNoise}_{\time}^{\np}\big)$
	\nllabel{alg:ienkf:filter:update}
      }
}   set $\vectheta_{\Nit}=\frac{1}{\Np}\sum_{\np=1}^{\Np}\vecTheta^{F,\Nit,\np}_{\Time}$
    \;
  \KwOut{
    Monte Carlo maximum likelihood estimate, $\vectheta_{\Nit}$.
  }
  \KwCplx{$\bigO(\Nit\Np\Unit\Time)$}
\end{algorithm}

\Cref{alg:ienkf} describes \code{enkf}, an implementation of the iterated ensemble Kalman filter ({\iEnKF}) which extends the IF2 approach for parameter estimation by replacing a particle filter with an ensemble Kalman filter.
The pseudocode uses the free index notation described in \cref{sec:igirf}.
An {\iEnKF} algorithm was demonstrated by \citet{li20}.
Alternative inference approaches via EnKF include using the EnKF likelihood within Markov chain Monte Carlo \citep{katzfuss20}.

{\iEnKF} uses the data to update the estimates of the latent state and parameters via a linear combination of the values in the ensemble (Line~\ref{alg:ienkf:filter:update}).
This causes difficulty estimating parameters which affect the spread of the ensemble distribution but not its center.
In particular, it can fail to estimate variance parameters.
For example, consider estimation of the measurement variance in the correlated Gaussian random walk model, \code{bm10}, with other parameters known.
In this case, the error vector $\vec{r}^j_n$ in Line~\ref{alg:ienkf:r} has zero expectation for any value of the measurement variance, $\tau$.
Thus, the increment to the parameter estimate in Line~\ref{alg:ienkf:filter:update} also has zero expectation for any value of $\tau$, and IEnKF fails.
By contrast, for the geometric Brownian motion model generated by \code{gbm()} corresponding to an exponentiation of this correlated Gaussian random walk model, {\iEnKF} can estimate $\tau$ because higher values of $\tau$ lead to higher expected values of $\vec{\hat{Y}}^j_n$ (Line~\ref{alg:ienkf:forecast:ensemble}).
In this case, if the average of the forecast ensemble is different from the observed data, the estimate of $\tau$ gets updated to reduce this discrepancy.
In summary, EnKF and IEnKF are numerically convenient algorithms, but care is needed to check whether they are suitable when developing a new model.

\subsection{Iterated block particle filter for parameter estimation}
The success of \code{bpfilter} on a variety of spatiotemporal models \citep{ionides23} raises the question of how to extend a block particle filter for parameter estimation.
An iterated filtering algorithm accommodating the structure of the block particle filter was proposed by \citet{ionides22}.
This generalizes the algorithm of \citet{ning23-ibpf} which addresses the special case where all parameters are unit-specific.
These algorithms are implemented by \code{ibpf}, which requires a \code{spatPomp} model with the property that estimated parameters can be perturbed across units as well as through time.
Therefore, any estimated parameter (whether shared or unit-specific) must be coded as a unit-specific parameter in order to apply this method.
The spatiotemporal perturbations are used only as an optimization tool for model parameters which are fixed though time and space (for shared parameters) or just through time (for unit-specific parameters).
The algorithm uses decreasing perturbation magnitudes so that the perturbed model approaches the fixed parameter model as the optimization proceeds.

An example model compatible with \code{ibpf} is constructed by the \code{he10()} function.
This builds a measles model similar to the \code{measles()} example discussed in \cref{sec:measles}, with the difference that the user can select which parameters are unit-specific. 
For further discussion of \code{ibpf}, and related questions about selecting shared versus unit-specific parameters, we refer the reader to \citet{ionides22}.
A separate tutorial guide provides additional detail on the use of \code{ibpf} \citep{ning23-tutorial}.

\subsection[Iterated UBF]{Iterated UBF for parameter estimation}
\label{sec:iubf}
\begin{algorithm}[p]
  \caption{\code{iubf(P,}\code{params{\argequals}$\vectheta_0$,}\code{Nubf{\argequals}$\Nit$,}\code{Nparam{\argequals}$\ntheta$,}\code{Nrep\_per\_param{\argequals}$\Rep$,}\code{nbhd=$B_{\unit,\time}$,}
     \code{prop=$p$,}\code{cooling.fraction.50{\argequals}$a$,}\code{rw.sd{\argequals}$\sigma_{0:\Time,1:\Thetadim}$)},
     using notation from \cref{tab:notation} where \code{P} is a
     `\code{\spatPomp}' object equipped with
     \code{rprocess},
     \code{dunit\_measure},
    \code{rinit},
    \code{obs} and \code{coef}.
  }\label{alg:iubf}
  \KwIn{
    simulator for $f_{\vec{X}_{\time}|\vec{X}_{\time-1}}(\vec{x}_{\time}\given \vec{x}_{\time-1}\giventh\vectheta)$ and $f_{\vec{X}_0}(\vec{x}_0\giventh\vectheta)$;
    evaluator for $f_{{Y}_{\unit,\time}|{X}_{\unit,\time}}({y}_{\unit,\time}\given {x}_{\unit,\time}\giventh\vectheta)$;
    data, $\data{\vec{y}}_{1:\Time}$;
    starting parameter, $\vectheta_0$;
    number of parameter vectors, $\ntheta$;
    number of replicates per parameter, $\Rep$;
    neighborhood structure, $B_{\unit,\time}$;
    number of iterations, $\Nit$;
    resampling proportion, $p$;
    random walk intensities, $\sigma_{0:\Time,1:\Thetadim}$;
    cooling fraction in 50 iterations, $a$.
  }
  \KwIndices{free indices are implicit `for' loops, calculated for $\rep \text{ in } \seq{1}{\Rep}$, $k$ in $\seq{1}{\ntheta}$, $\unit$ and $\altUnit$ in $\seq{1}{\Unit}$, $\thetadim$ and $\thetadim^\prime$ in $\seq{1}{\Thetadim}$.}
    \BlankLine
initialize parameters,
    $\vecTheta^{F,0,k}_{\Time} = \vectheta_{0}$
\;
\For{$\nit\ \mathrm{in}\ \seq{1}{\Nit}$}{
  initialize parameters,
      $\Theta^{F,\nit,k}_{0}=
        \Theta^{F,\nit-1,k}_{\Time}$   
  \;
  initialize filter particles,
  $\vec{X}_{0}^{F,\nit,k,\rep}\sim {f}_{\vec{X}_{0}}
    \left(\mydot\giventh{\vecTheta^{F,\nit,k}_{0}}\right)$
  \;
  \For{$\time\ \mathrm{in} \ \seq{1}{\Time}$}{
     perturb parameters,
     $\Theta_{\time}^{P,\nit,k,\rep} \sim  \normal \big(
       \Theta^{F,\nit,k}_{\time-1} \, \, , \,
       a^{2\nit/50} \, \Sigma_{\time}
       \big)$,
     where
     $\big[\Sigma_{\time}\big]_{\thetadim,\thetadim^\prime} =
       \sigma_{\time,\thetadim}^2 \mathbbm{1}_{\thetadim=\thetadim^\prime}$
  \;
  \nllabel{alg:iabf:for:n}
  proposals,
  $\vec{X}_{\time}^{\IP,\nit,k,\rep} \sim
    f_{\vec{X}_{\time}|\vec{X}_{\time-1}}
    \big( \mydot \given \vec{X}^{F,\nit,k,\rep}_{\time-1}
    \giventh \Theta_{\time}^{P,\nit,k,\rep} \big)
  $
  \nllabel{alg:iabf:adapted:proposals}
  \;
  unit measurement density,
  $\unitWeight_{\unit,\time}^{k,\rep} =
    f_{Y_{\unit,\time}|X_{\unit,\time}}
    \big (\data{y}_{\unit,\time}\given X^{\IP,\nit,k,\rep}_{\unit,\time} \giventh \Theta_{\time}^{P,\nit,k,\rep}\big)$
  \;
  local prediction weights,
  $w^{\LCP,k,\rep}_{\unit,\time}= \displaystyle
       \prod_{(\altUnit,\altTime)\in B_{\unit,\time}}
         \hspace{-2mm}
        \unitWeight_{\altUnit,\altTime}^{k,\rep}$
  \nllabel{alg:iabf:lp:weights} \nllabel{alg:iabf:end:n}
  \;
  parameter log-likelihoods, 
  $\displaystyle r^{k}_{\time}= \sum_{\unit=1}^{\Unit}\mathrm{log} \Bigg(
    \frac{
      \sum_{\rep=1}^{\Rep}\unitWeight_{\unit,\time}^{k,\rep} \, \,
      w^{\LCP,k,\rep}_{\unit,\time}
    }{
      \sum_{\altRep=1}^{\Rep}
      w^{\LCP,k,\altRep}_{\unit,\time}
    }\Bigg)
  $
  for $k$ in $\seq{1}{\ntheta}$, 
  \nllabel{alg:iabf:rep:weights}
  \;
  Select the highest $p\ntheta$ weights:
  find $s$ with
  $\{s(1),\dots,s(p\ntheta)\}=
    \big\{k: \sum_{\tilde{k}=1}^{\ntheta}{\mathbf{1}} 
    \{r^{\tilde{k}} > r^{k} \} < p \ntheta
    \big\}$
  \;
  Make $1/p$ copies of successful parameters,
  $\Theta^{F,\nit,k}_{\time}=\Theta^{F,\nit, s(\lceil p k\rceil)}_{\time}$
  for $k$ in $\seq{1}{\ntheta}$
  \;
  Set  $\vec{X}^{F,\nit,k,\rep}_{\time} = \vec{X}_{\time}^{\IP,\nit,s(\lceil p k\rceil),\rep}$
  \;
  }
}
  \KwOut{
    Iterated UBF parameter swarm: $\Theta^{F, \Nit, 1:\ntheta}_{\Time}$ \; Monte Carlo maximum likelihood estimate: $\frac{1}{\ntheta} \sum_{k=1}^{\ntheta}\Theta^{F, \Nit, 1:\ntheta}_{\Time}$.
  }
  \KwCplx{$\bigO(\Nit\ntheta\Rep\Unit\Time)$}

\end{algorithm}

Algorithm \ref{alg:iubf} describes the \code{iubf} function which carries out parameter estimation by iterating an unadapted bagged filter (UBF) with perturbed parameters.
Note that UBF is the special case of ABF with $\Np=1$.
UBF and IUBF were found to be effective on the measles model \citep{ionides23} and \code{iubf} was developed with this application in mind.
This algorithm makes with $\ntheta$ copies of the parameter set, and iteratively perturbs the parameter set while evaluate a conditional likelihood at each observation time using UBF.
In each observation time, IUBF selects perturbed parameter sets yielding the top $p$ quantile of the likelihoods.

\subsection{Inference algorithms inherited from pomp}
\label{sec:otherinference}
Objects of \class{spatPomp} inherit methods for inference from \class{pomp} objects implemented in the \pkg{pomp} package.
As discussed earlier, the IF2 algorithm \citep{ionides15} has been used for maximum likelihood parameter estimation in numerous applications.
IF2 can be used to check the capabilities of newer and more scalable inference methods on smaller examples for which it is known to be effective.
Extensions for Bayesian inference of the currently implemented high-dimensional particle filter methods (GIRF, ABF, EnKF, BPF) are not yet available.
Bayesian inference is available in \pkg{spatPomp} using the approximate Bayesian computing (ABC) method inherited from \pkg{pomp}, \code{abc()}.
ABC has previously been used for spatiotemporal inference \citep{brown2018approximate} and can also serve as a baseline method.
However, ABC is a feature-based method that may lose substantial information compared to full-information methods that work with the likelihood function rather than a summary statistic \citep{fasiolo16}.

%%%%%%%%%%%%%%%%%%%%%%%%%%%%%%%%%%%%%%%%%%%%%%%%%%%%%%%%%%%%

\section[Demonstration of data analysis tools]{Demonstrating data analysis tools on a toy model}
\label{sec:examples}
\label{sec:brownian}

We illustrate key capabilities of \pkg{spatPomp} using the \code{bm10} model for correlated Brownian motion introduced in Sec.\ref{section:included}.
This allows us to demonstrate a data analysis in a simple context where we can compare results with a standard particle filter as well as validate all methods against the exact solutions which are analytically available.
To define the model mathematically, consider spatial units $1,\dots,\Unit$ located evenly around a circle, where $\dist(\unit,\altUnit)$ is the circle distance,
\[
\dist(\unit,\altUnit)
= \min\big(|\unit-\altUnit|, |\unit-\altUnit+\Unit|, |\unit-\altUnit-\Unit|\big).
\]
The latent process is a $\Unit$-dimensional Brownian motion $\vec{X}(t)$ having correlation that decays with distance.
Specifically,
\[
dX_\unit(t)= \sum_{\altUnit=1}^{\Unit} \rho^{\dist(\unit,\altUnit)} \, dW_{\altUnit}(t),
\]
where $W_{1}(t),\dots,W_{\Unit}(t)$ are independent Brownian motions with infinitesimal variance $\sigma^2$, and $|\rho| <1$.
Using the notation in Section \ref{sec:background}, we suppose our measurement model for discrete-time observations of the latent process is
\[
Y_{\unit,\time}=X_{\unit,\time} + \eta_{\unit,\time}
\]
where  $\eta_{\unit,\time}\overset{\text{iid}}{\sim} \normal(0,\tau^2)$.
The model is completed by providing the initial conditions, $\{X_\unit(0), \unit \in 1:\Unit \}$, which are specified as parameters.
The parameters for \code{bm10} are
\begin{knitrout}
\definecolor{shadecolor}{rgb}{1, 1, 1}\color{fgcolor}\begin{kframe}
\begin{verbatim}
R> coef(bm10)
  rho sigma   tau  X1_0  X2_0  X3_0  X4_0  X5_0  X6_0  X7_0  X8_0 
  0.4   1.0   1.0   0.0   0.0   0.0   0.0   0.0   0.0   0.0   0.0 
 X9_0 X10_0 
  0.0   0.0 
\end{verbatim}
\end{kframe}
\end{knitrout}

\subsection{Computing the likelihood}
\label{sec:loglik}

The \code{bm10} object contains all necessary model components for likelihood evaluation using the four algorithms described in \Cref{sec:filtering}.
For example,
\begin{knitrout}
\definecolor{shadecolor}{rgb}{1, 1, 1}\color{fgcolor}\begin{kframe}
\begin{verbatim}
R> girf(bm10,Np=500,Nguide=50,Ninter=5,lookahead=1)
R> bpfilter(bm10, Np=2000, block_size=2)
R> enkf(bm10, Np=2000)
\end{verbatim}
\end{kframe}
\end{knitrout}
\noindent This generates objects of class `\code{girfd\_spatPomp}', `\code{bpfiltered\_spatPomp}` and `\code{enkfd\_spatPomp}' respectively.
A \code{plot} method provides diagnostics, and the resulting log-likelihood estimate is extracted by \code{logLik}.

\begin{knitrout}
\definecolor{shadecolor}{rgb}{1, 1, 1}\color{fgcolor}\begin{figure}

{\centering \includegraphics[width=0.75\linewidth]{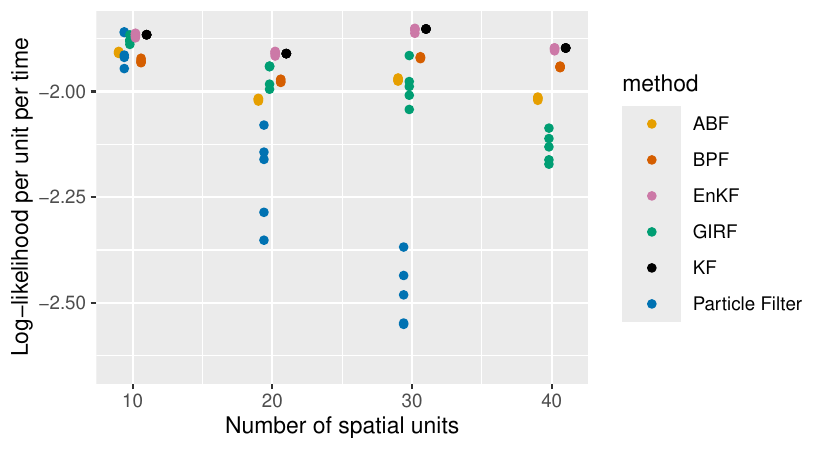} 

}

\caption[Log-likelihood estimates for 5 replications of ABF, BPF, EnKF, GIRF and particle filter on correlated Brownian motions of various dimensions]{Log-likelihood estimates for 5 replications of ABF, BPF, EnKF, GIRF and particle filter on correlated Brownian motions of various dimensions. The Kalman filter (KF) provides the exact likelihood in this case.}\label{fig:bm_lik_plot}
\end{figure}

\end{knitrout}

Even for methods designed to be scalable, Monte Carlo variance can be expected to grow with the size of the dataset, and approximations used to enhance scalability may result in bias.
\Cref{fig:bm_lik_plot} investigates how the accuracy of the likelihood estimate scales with $U$ for the \code{bm} model.
Since \class{spatPomp} inherits from \class{pomp}, we can compare \pkg{spatPomp} methods against the \code{pfilter} algorithm from \pkg{pomp}.
We see that the performance of \code{pfilter} rapidly degrades as dimension increases, whereas the \pkg{spatPomp} methods scale better.
On this Gaussian problem, the exact likelihood is available via the Kalman filter, and EnKF is almost exact since the Gaussian approximation used to construct its update rule is correct.

Computing resources used by each algorithm for \cref{fig:bm_lik_plot} are given in \cref{tab:bm_table_df}.
Each algorithm was allowed to use 10 central processing unit (CPU) cores to evaluate all the likelihoods and the algorithmic settings were fixed as shown in the table.
CPU time is not necessarily the only relevant consideration, for example, when applying \code{pfilter} with a large number of units and a complex model, memory constraints rather than CPU requirements may limit the practical number of particles.
By contrast, ABF has a high CPU requirement but it parallelizes easily to take advantage of distributed resources. 

The time-complexity of GIRF is quadratic in $\Unit$, due to the intermediate time step loop shown in the pseudocode in \cref{sec:girf}, whereas the other algorithms scale linearly with $\Unit$ for a fixed algorithmic setting.
However, a positive feature of GIRF is that it shares with PF the property that it targets the exact likelihood, i.e., it is consistent for the exact log-likelihood as the number of particles grows and the Monte Carlo variance approaches zero.
GIRF may be a practical algorithm when the number of units prohibits PF but permits effective use of GIRF.
EnKF and BPF generally run the quickest and require the least memory.
However, the Gaussian and independent blocks assumptions, respectively, of the two algorithms must be reasonable to obtain likelihood estimates with low bias.
On a new problem, it is advantageous to compare various algorithms to reveal unexpected limitations of the different approximations inherent in each algorithm.

\begin{knitrout}
\definecolor{shadecolor}{rgb}{1, 1, 1}\color{fgcolor}\begin{table}

\caption{\label{tab:bm_table_df}Comparison of computational resources of the filtering algorithms}
\centering
\begin{tabular}[t]{lccccc}
\toprule
\makecell[l]{Method \\ { }} & \makecell[l]{Resources \\ (core-minutes)} & \makecell[l]{Particles \\ (per replicate)} & \makecell[l]{Replicates \\ { }} & \makecell[l]{Guide \\ particles} & \makecell[l]{Lookahead \\ { }}\\
\midrule
Particle Filter & 1.00 & 2000 & - & - & -\\
ABF & 48 & 100 & 500 & - & -\\
GIRF & 12 & 500 & - & 50 & 1\\
EnKF & 1.1 & 2000 & - & - & -\\
BPF & 1.6 & 2000 & - & - & -\\
\bottomrule
\end{tabular}
\end{table}

\end{knitrout}

\subsection{Parameter inference}
\label{sec:bm:inference}
The correlated Brownian motions example also serves to illustrate parameter inference using IGIRF.
Suppose we have data from the correlated 10-dimensional Brownian motions model discussed above.
We consider estimation of the parameters $\sigma$, $\tau$ and $\rho$ when that the initial conditions, $\{{X}_{\unit}(0), \unit \in 1\mycolon\Unit \}$, are known to be zero.
We demonstrate a search started at
\begin{knitrout}
\definecolor{shadecolor}{rgb}{1, 1, 1}\color{fgcolor}\begin{kframe}
\begin{verbatim}
R> start_params <- c(rho = 0.8, sigma = 0.4, tau = 0.2,
+    X1_0 = 0, X2_0 = 0, X3_0 = 0, X4_0 = 0, X5_0 = 0,
+    X6_0 = 0, X7_0 = 0, X8_0 = 0, X9_0 = 0, X10_0 = 0)
\end{verbatim}
\end{kframe}
\end{knitrout}
We start with a test of \code{igirf}, estimating the parameters $\rho$, $\sigma$ and $\tau$ but not the initial value parameters.
We use a computational intensity variable, \code{i}, to switch between algorithmic parameter settings.
For debugging, testing and code development we use \code{i=1}.
For a final version of the manuscript, we use \code{i=2}.
\begin{knitrout}
\definecolor{shadecolor}{rgb}{1, 1, 1}\color{fgcolor}\begin{kframe}
\begin{verbatim}
R> i <- 2
R> ig1 <- igirf(
+    bm10,
+    params=start_params,
+    Ngirf=switch(i,2,50),
+    Np=switch(i,10,1000),
+    Ninter=switch(i,2,5),
+    lookahead=1,
+    Nguide=switch(i,5,50),
+    rw.sd=rw_sd(rho=0.02,sigma=0.02,tau=0.02),
+    cooling.type = "geometric",
+    cooling.fraction.50=0.5
+  )
\end{verbatim}
\end{kframe}
\end{knitrout}

\code{ig1} is an object of \class{igirfd\_spatpomp} which inherits from \class{girfd\_spatpomp}.
A useful diagnostic of the parameter search is a plot of the change of the parameter estimates during the course of an \code{igirf()} run.
Each iteration within an \code{igirf} run provides a parameter estimate and a likelihood evaluation at that estimate.
The \code{plot} method for a \class{igirfd\_spatPomp} object shows the convergence record of parameter estimates and their likelihood evaluations.
As shown in \cref{fig:bm_igirf_convergence2}, this \code{igirf} search has allowed us to explore the parameter space and climb significantly up the likelihood surface to within a small neighborhood of the maximum likelihood.
The search took 12.7 minutes on one CPU core for this example with 10 spatial units.
Investigation of larger models may require multiple searches of the parameter space, started at various points, implemented using parallel runs of \code{igirf()}.

The log-likelihood plotted in \cref{fig:bm_igirf_convergence2}, and that computed by \code{logLik(ig1)}, correspond to the perturbed model.
These should be recomputed to obtain a better estimate for the unperturbed model.
Different likelihood evaluation methods can be applied, as shown in \cref{sec:loglik}, to investigate their comparative strengths and weaknesses.
For \code{bm10}. the model is linear and Gaussian and so the maximum likelihood estimate of our model and the likelihood at this estimate can be found numerically using the Kalman filter.
The maximum log-likelihood is -373.0, whereas the likelihood obtained by \code{ig1} is -374.2.
This shortfall is a reminder that Monte Carlo optimization algorithms should usually be replicated, and should be used with inference methodology that accommodates Monte Carlo error, as discussed in \cref{sec:mcap}.

\begin{knitrout}
\definecolor{shadecolor}{rgb}{1, 1, 1}\color{fgcolor}\begin{figure}

{\centering \includegraphics[width=0.65\linewidth]{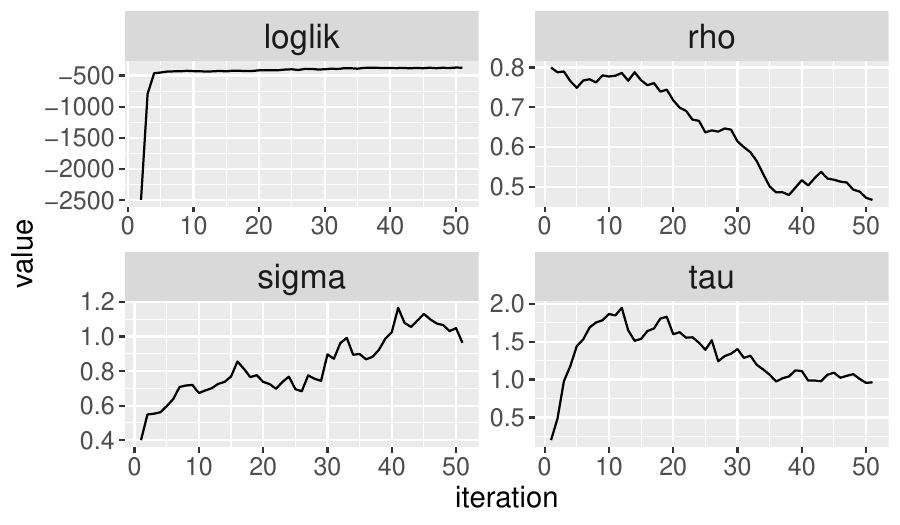} 

}

\caption{The output of the \code{plot()} method on the object of \class{igirfd\_spatPomp} that encodes our model for correlated Brownian motions produces convergence traces for $\rho$, $\sigma$ and $\tau$, and the corresponding log-likelihoods. Over 50 iterations \code{igirf()} has allowed us to get within a neighborhood of the maximum likelihood.}\label{fig:bm_igirf_convergence2}
\end{figure}

\end{knitrout}

\subsection{Monte Carlo profiles}
\label{sec:mcap}

Proper interpretation of a parameter estimate requires understanding its uncertainty.
Here, we construct a profile likelihood 95\% confidence interval for the coupling parameter, $\rho$, in the \code{bm10} model.
This entails calculation of the maximized likelihood over all parameters excluding $\rho$, for a range of fixed values of $\rho$.
We use Monte Carlo adjusted profile (MCAP) methodology to accommodate Monte Carlo error in maximization and likelihood evaluation \citep{ionides17,ning21}.

In practice, we carry out multiple searches for each value of $\rho$, with other parameters drawn at random from a specified hyperbox.
We build this box on a transformed scale suitable for optimization, taking advantage of the \code{partrans} method.
It is generally convenient to optimize non-negative parameters on a log scale and $(0,1)$ valued parameter on a logit scale.
We set this up using the \pkg{pomp} function \code{profile\_design}, taking advantage of the \code{partrans} method defined by the \code{partrans} argument to \code{spatPomp}, defined here as
\begin{knitrout}
\definecolor{shadecolor}{rgb}{1, 1, 1}\color{fgcolor}\begin{kframe}
\begin{verbatim}
R> bm10 <- spatPomp(bm10,
+    partrans = parameter_trans(log = c("sigma", "tau"), logit = c("rho")),
+    paramnames = c("sigma","tau","rho")
+  )
\end{verbatim}
\end{kframe}
\end{knitrout}
This provides access to the \code{partrans} method which we use when constructing starting points for the search:
\begin{knitrout}
\definecolor{shadecolor}{rgb}{1, 1, 1}\color{fgcolor}\begin{kframe}
\begin{verbatim}
R> theta_lo_trans <- partrans(bm10,coef(bm10),dir="toEst") - log(2)
R> theta_hi_trans <- partrans(bm10,coef(bm10),dir="toEst") + log(2)
R> profile_design(
+    rho=seq(from=0.2,to=0.6,length=10),
+    lower=partrans(bm10,theta_lo_trans,dir="fromEst"),
+    upper=partrans(bm10,theta_hi_trans,dir="fromEst"),
+    nprof=switch(i,2,10)
+  ) -> pd
\end{verbatim}
\end{kframe}
\end{knitrout}
The argument \code{nprof} sets the number of searches, each started at a random starting point, for each value of the profiled parameter, \code{rho}.
We can apply any of the methods of \cref{sec:inference} for likelihood maximization and any of the methods of \cref{sec:filtering} for likelihood evaluation.
Experimentation is recommended---here, we demonstrate using \code{igirf} and \code{enkf}.
We run parallel searches using \code{foreach} and \code{\%dopar\%} from the \code{foreach} package \citep{R:foreach}  and collecting all the results together using \code{bind\_rows} from \code{dplyr} \citep{R:dplyr}.
Multiple log-likelihood evaluations are carried out on the parameter estimate resulting from each search, averaged using \code{logmeanexp} which also provides a standard error.
\begin{knitrout}
\definecolor{shadecolor}{rgb}{1, 1, 1}\color{fgcolor}\begin{kframe}
\begin{verbatim}
R> foreach (p=iter(pd,"row"),.combine=dplyr::bind_rows) %dopar% {
+    library(spatPomp)
+    ig2 <- igirf(ig1,params=p,rw.sd=rw_sd(sigma=0.02,tau=0.02))
+    ef <- replicate(switch(i,2,10),enkf(ig2,Np=switch(i,50,2000)))
+    ll <- sapply(ef,logLik)
+    ll <- logmeanexp(ll,se=TRUE)
+    data.frame(as.list(coef(ig2)),loglik=ll[1],loglik.se=ll[2])
+  } -> rho_prof
\end{verbatim}
\end{kframe}
\end{knitrout}

Above, calling \code{igirf} on \code{ig1} imports all the previous algorithmic settings except for those that we explicitly modify.
Each row of \code{rho\_prof} now contains a parameter estimate its log-likelihood, with $\rho$ values fixed along a grid.
The MCAP 95\% confidence interval constructed by \code{mcap} uses \code{loess} to obtain a smoothed estimate of the profile likelihood function and then determines a confidence interval using by a cutoff based on the delta method applied to a local quadratic regression.
This cutoff is typically slightly larger than the asymptotic 1.92 cutoff for a standard profile likelihood confidence interval constructed assuming error-free likelihood maximization and evaluation.
\begin{knitrout}
\definecolor{shadecolor}{rgb}{1, 1, 1}\color{fgcolor}\begin{kframe}
\begin{verbatim}
R> rho_mcap <- mcap(rho_prof[,"loglik"],parameter=rho_prof[,"rho"])
R> rho_mcap$ci
[1] 0.2640641 0.5075075
\end{verbatim}
\end{kframe}
\end{knitrout}
Note that the data in \code{bm10} are generated from a model with $\rho=0.4$.

\section[Measles model]{A spatiotemporal model of measles transmission}
\label{sec:measles}
A \pkg{spatPomp} data analysis may consist of the following major steps: (i) obtain data, postulate a class of models that could have generated the data and bring these two pieces together via a call to \texttt{spatPomp()};
(ii) employ the tools of likelihood-based inference, evaluating the likelihood at specific parameter sets, maximizing likelihoods under the postulated class of models, constructing Monte Carlo adjusted confidence intervals, or performing likelihood ratio hypothesis tests of nested models; 
(iii) criticize the model by comparing simulations to data, or by considering rival models.
In this section, we focus on step (i), showing how to bring data and models together via a metapopulation compartment model for measles dynamics in the 6 largest cities in England in the pre-vaccine era.
Tools for step (ii) have been covered in Sections~\ref{sec:filtering} and~\ref{sec:inference}.
Step (iii) benefits from the flexibility of the large model class supported by \pkg{spatPomp}.

Compartment models for population dynamics partition the population into categories called \textit{compartments}.
Individuals may move between compartments, and the rate of flow of individuals between a pair of compartments may depend on the number of individuals in other compartments.
Compartment models have widespread scientific applications, especially in the biological and health sciences \citep{breto09}.
Spatiotemporal compartment models can be called patch models or metapopulation models in an ecological context, where each spatial unit is called a patch or a sub-population.
We present a spatiotemporal model for disease transmission dynamics of measles within and between multiple cities, based on the model of \citet{park20} which adds spatial interaction to the compartment model presented by \citet{he10}.
We write equations for a SpatPOMP model constructed by the \code{measles()} and \code{he10} functions in \pkg{\spatPomp} construct.
This illustrates how \pkg{spatPomp} can accommodate various model features that may be relevant for a successful statistical description of epidemiological metapopulation dynamics.
We then demonstrate explicitly how to construct a simplified version of this model.

The \code{measles()} and \code{he10()} models are similar, but differ in details.
For \code{measles()}, the data consist of biweekly counts and parameters are all shared between units, matching the analysis of \citet{park20} and \citet{ionides23}.
For \code{he10()}, data are weekly and parameters can be shared or unit-specific, matching the analysis of \citet{he10} and \citet{ionides22}.
We write the model for the general case where all parameters are unit-specific, noting that it is a relevant data analysis question to determine when parameter dependence on $\unit$ can be omitted.

\subsection{Mathematical model for the latent process}
\label{sec:measles-math-model}
We first define the model mathematically, starting with a description of the coupling, corresponding here to travel between cities.
Let $v_{\unit\altUnit}$ denote the number of travelers from city $\unit$ to $\altUnit$.
Here, $v_{\unit\altUnit}$ follows the gravity model of \citet{xia04}, with
$v_{\unit\altUnit} = \gravity_{\unit} V_{\unit\altUnit}$ where $\gravity_{\unit}$ is called a gravitation parameter and
\[
V_{\unit\altUnit} =  \frac{\popTimeAvg_\unit \cdot \popTimeAvg_{\altUnit}}{\dist(\unit,\altUnit)} \times \frac{\;\overline{\dist}\;}{\;\, \overline{\popTimeAvg}^2 \, \;}.
\]
Here, $\dist(\unit,\altUnit)$ denotes the distance between city $\unit$ and city $\altUnit$, $\popTimeAvg_{\unit}$ is the average across time of the census population $P_{\unit}(t)$ for city $\unit$, $\overline{\popTimeAvg}$ is the average of $\popTimeAvg_{\unit}$ across cities, and $\overline{\dist}$ is the average of $\dist(\unit,\altUnit)$ across pairs of cities.

The measles model divides the population of each city into susceptible, $S$, exposed, $E$, infectious, $I$, and recovered/removed, $R$, compartments.
The number of individuals in each compartment for city $\unit$ at time $t$ are denoted by $S_\unit(t)$, $E_\unit(t)$, $I_\unit(t)$, and $R_\unit(t)$.
The latent state is $\vec{X}(t)=\big(X_1(t),\dots,X_{\Unit}(t)\big)$ with $X_{\unit}(t) = \big(S_{\unit}(t), E_{\unit}(t),I_{\unit}(t),R_{\unit}(t)\big)$.
The dynamics of the latent state can be written in terms of flows between compartments, together with flows into and out of the system, as follows:
\[
\left.
\begin{array}{lllllll}
\displaystyle dS_\unit(t) &=& dN_{BS,\unit}(t) &-& dN_{SE,\unit}(t) &-& dN_{SD,\unit}(t) \\
\displaystyle dE_\unit(t) &=& dN_{SE,\unit}(t) &-& dN_{EI,\unit}(t) &-& dN_{ED,\unit}(t) \\
\displaystyle dI_\unit(t) &=& dN_{EI,\unit}(t) &-& dN_{IR,\unit}(t) &-& dN_{ID,\unit}(t)
\end{array}\qquad\right\}\quad \mbox{for $\unit=1,\dots, \Unit$.}
\]
Here, $N_{SE,\unit}(t)$, $N_{EI,\unit}(t)$, and $N_{IR,\unit}(t)$ are counting process corresponding to the cumulative number of individuals transitioning between the compartments identified by the subscripts.
The recruitment of susceptible individuals into city $\unit$ is denoted by the counting process $N_{BS,\unit}(t)$, primarily modeling births.
Each compartment also has an outflow, written as a transition to $D$, primarily representing death, which occurs at a constant per-capita rate $\mu$.
The number of recovered individuals $R_\unit(t)$ in city $\unit$ is defined implicitly from $P_{\unit}(t)=S_{\unit}(t)+E_{\unit}(t)+I_{\unit}(t)+R_{\unit}(t)$.
$R_{\unit}(t)$ plays no direct role in the dynamics, beyond accounting for individuals not in any of the other classes.

To define the Markov model, we specify a rate for each counting process.
Thus, $\mu_{EI,\unit}$ is the rate at which an individual in $E$ progresses to $I$ in city $\unit$, and $1/\mu_{EI,\unit}$ is called the mean disease latency.
Similarly, $1/\mu_{IR,\unit}$ is the mean infectious period.
The mortality rates are fixed at $\mu_{SD,\unit}=\mu_{ED,\unit}=\mu_{ID,\unit}=\mu_{RD,\unit}=\mu_D$, with life expectancy $1/\mu_D = 50\,\mathrm{yr}$.
The rate of recruitment of susceptible individuals, $\mu_{BS,\unit}(t)$, is treated as a covariate defined in terms of the birth rate, $b_{\unit}(t)$, known from public records.
Specifically,
\begin{equation}\nonumber
\mu_{BS,\unit}(t) = b_{\unit}(t-t_b)\big[(1-\cohort_{\unit})+\cohort_{\unit}\,\delta(t-t_a)\big],
\end{equation}
where $t_a=\lfloor t \rfloor + 252/365$ is the school admission date for the year containing $t$, $t_b=4\,\mathrm{yr}$ is a fixed delay between birth and entry into the high-transmission community, $\cohort_{\unit}$ is a fraction of the births which join the $S$ on their first day of school, and $\delta$ is the Dirac delta function.
All rates other than $\mu_{BS,\unit}$ are defined per capita.
The disease transmission rate, $\mu_{SE,\unit}$, is parameterized as
\begin{equation*} \label{eq:muSE}
\mu_{SE,\unit}(t)= \betaBar_{\unit} \, \measlesSeasonality_{\unit}(t)
  \left[
    \left(\frac{I_{\unit}(t)+\measlesImmigration_{\unit}}{P_{\unit}(t)}\right)^{\alpha_{\unit}} +
    \sum_{\altUnit\neq\unit}
      \frac{v_{\unit\altUnit}}{P_{\unit}(t)} \left\{
        \left(\frac{I_{\altUnit}(t)}{P_{\altUnit}(t)}\right)^{\alpha_{\altUnit}} -
	\left(\frac{I_{\unit}(t)}{P_{\unit}(t)} \right)^{\alpha_{\unit}}
      \right\}
  \right]
   \frac{d\Gamma_{SE,\unit}}{dt},
\end{equation*}
where the mean transmission rate, $\betaBar_{\unit}$, is parameterized as $\betaBar_{\unit}=\Rzero{,\unit}(\mu_{IR,\unit}+\mu_D)$ with $\Rzero{,\unit}$ being the basic reproduction rate; 
$\measlesSeasonality_{\unit}(t)$ is a periodic step function taking value $(1-\measlesAmplitude_{\unit})$ during school vacations and $(1+0.381\, \measlesAmplitude_{\unit})$ during school terms, defined so that the average value of $\measlesSeasonality_{\unit}(t)$ is 1;
$\alpha_{\unit}$ is an exponent describing non-homogeneous mixing of individuals;
$\measlesImmigration_{\unit}$ describes infected individuals arriving from outside the study population;
the multiplicative white noise $d\Gamma_{SE,\unit}/dt$ is a derivative of a gamma process $\Gamma_{SE,\unit}(t)$ having independent gamma distributed increments with $\E[\Gamma_{SE,\unit}(t)]=t$ and $\var[\Gamma_{SE,\unit}(t)]=\sigma^2_{SE,\unit}t$, where $\sigma^2_{SE,\unit}$ is the infinitesimal variance of the noise.
The formal meaning of $d\Gamma_{SE,\unit}/dt$ as white noise on the rate of a Markov chain was developed by \citet{breto09} and \citet{breto11}.
In brief, an Euler numerical solution depends on the rate function integrated over a small time interval of length $\Delta t$.
The integrated noise process is an increment of the gamma process, and these increments are independent gamma random variables.
The continuous time Markov chain corresponding to this noisy rate is the limit of the Euler solutions as $\Delta t \rightarrow 0$, and so these Euler solutions provide a practical approach to working with the model.
A single Euler step is defined via the Csnippet for \code{rprocess}, below.
This code involves use of the \code{reulermultinom} function, which is the \proglang{C} interface to the \proglang{R} function \code{reulermultinom} provided by \pkg{pomp}.
It keeps track of all the rates for possible departures from a compartment.
The gamma white noise in these rates is added using the \code{rgammawn} function, which is also defined by \pkg{pomp} in both \proglang{C} and \proglang{R}.

Multiplicative white noise provides a way to model over-dispersion, a phenomenon where data variability is larger than can be explained by binomial or Poisson approximations.
Over-dispersion on a multiplicative scale is also called environmental stochasticity, or logarithmic noise, or extra-demographic stochasticity.
Over-dispersion is well established for generalized linear models \citep{mccullagh89} and has become increasingly apparent for compartment models as methods have become available to address it \cite{bjornstad01,he10,stocks20}.

Initial conditions for the latent state process at a time $t_0$ are described in terms of initial value parameters, $S_{\unit,0}$, $E_{\unit,0}$ and $I_{\unit,0}$, defined as follows:
\begin{equation}\nonumber
\begin{array}{rclrcl}
S_{\unit}(t_0)&=&\mathrm{round}\big(S_{\unit,0} \, P_{\unit}(t_0)\big), &
  \hspace{5mm}E_{\unit}(t_0)&=&\mathrm{round}\big(E_{\unit,0} \,P_{\unit}(t_0)\big),
\\
I_{\unit}(t_0)&=&\mathrm{round}\big(I_{\unit,0}\, P_{\unit}(t_0)\big), &
  R_{\unit}(t_0)&=& P_{\unit}(t_0)-S_{\unit}(t_0) - E_{\unit}(t_0) - I_{\unit}(t_0).
\end{array}
\end{equation}

The observations for city $\unit$ are bi-weekly reports of new cases.
We model the total new cases in an interval by keeping track of transitions from $I$ to $R$, since we expect that identified cases will typically be isolated from susceptible individuals. 
Therefore, we introduce a new latent variable, defined at observation times as
\begin{equation*}
C_{\unit,n} = N_{IR,\unit}(t_n) - N_{IR,\unit}(t_{n-1}).
\end{equation*}
To work with $C_{\unit,n}$ in the context of a SpatPOMP model, we note that this variable has Markovian dynamics corresponding to a continuous time variable $C_{\unit}(t)$ satisfying $dC_{\unit}(t)=dN_{IR,\unit}(t)$ with the additional property that we set $C_{\unit}(t)=0$ immediately after an observation time.
To model the observation process, we define $Y_{\unit,n}$ as a normal approximation to an over-dispersed binomial sample of $C_{\unit,n}$ with reporting rate $\rho_{\unit}$. 
Specifically, conditional on $C_{\unit,n} = c_{\unit,n}$,
\begin{equation*}
Y_{\unit,\time} \sim \normal
  \big[
    \rho_{\unit}\,c_{\unit,n},\rho\,(1-\rho_{\unit})\,c_{\unit,n}+\measlesMeasurementOD^2\rho_{\unit}^2c_{\unit_n}^2
  \big],
\end{equation*}
where $\measlesMeasurementOD$ is a measurement overdispersion parameter.

\subsection{Construction of a measles spatPomp object}
\label{sec:measles:spatPomp}

We construction the model described in \cref{sec:measles-math-model} for the simplified situation where $\alpha_{\unit}=1$, $\measlesImmigration_{\unit}=0$ and $\cohort_{\unit}=0$.
All other parameters have shared value across units, except for the initial value parameters.
A complete \pkg{spatPomp} representation of the model is provided in the source code for \code{he10()}.

We use the bi-weekly measles case counts from $\Unit=6$ cities in England as reported by \citet{dalziel16}, provided in the object \code{measles\_cases}.
Each city has about 15 years (391 bi-weeks) of data, with no missing data.
The first three rows of this data are shown below, with the \code{year} column corresponding to the observation date in years.
\begin{knitrout}
\definecolor{shadecolor}{rgb}{1, 1, 1}\color{fgcolor}\begin{kframe}
\begin{verbatim}
     year       city cases
 1950.000     LONDON    96
 1950.000 BIRMINGHAM   179
 1950.000  LIVERPOOL   533
 1950.000 MANCHESTER    22
 1950.000      LEEDS    17
 1950.000  SHEFFIELD    48
 1950.038     LONDON    60
 1950.038 BIRMINGHAM   160
\end{verbatim}
\end{kframe}
\end{knitrout}
We can construct a \code{\spatPomp} object by supplying three minimal requirements in addition to our data above: the column names corresponding to the units labels (\code{`city'}) and observation times (\code{`year'}) and the time at which the latent dynamics are initialized. Here we set this to two weeks before the first recorded observations.
\begin{knitrout}
\definecolor{shadecolor}{rgb}{1, 1, 1}\color{fgcolor}\begin{kframe}
\begin{verbatim}
R> measles6 <- spatPomp(
+    data=measles_cases,
+    units='city',
+    times='year',
+    t0=min(measles_cases$year)-1/26
+  )
\end{verbatim}
\end{kframe}
\end{knitrout}
Internally, unit names are mapped to an index $1,\dots,\Unit$.
The number assigned to each unit can be checked by inspecting their position in \code{unit\_names(measles)}.
We proceed to collect together further model components, which we will add to \code{measles6} by a subsequent call to \code{spatPomp()}.
First, we suppose that we have covariate time series, consisting of census population, $P_{\unit}(t)$, and lagged birthrate, $b_{\unit}(t-t_b)$, in a \class{data.frame} object called \code{measles\_covar}.
The required format is similar to the \code{data} argument, though the times do not have to correspond to observation times since \pkg{spatPomp} will interpolate the covariates as needed.
\begin{knitrout}
\definecolor{shadecolor}{rgb}{1, 1, 1}\color{fgcolor}\begin{kframe}
\begin{verbatim}
 year       city lag_birthrate         P
 1950     LONDON      66318.99 3389306.0
 1950 BIRMINGHAM      22968.58 1117892.5
 1950  LIVERPOOL      18732.87  802064.9
\end{verbatim}
\end{kframe}
\end{knitrout}

We now move on to specifying our model components as Csnippets.
To get started, we define the movement matrix $\big(v_{\unit,\altUnit}\big)_{\unit,\altUnit \in 1\mycolon\Unit}$ as a global variable in \proglang{C} that will be accessible to all model components, via the \code{globals} argument to \code{spatPomp()}.
\begin{knitrout}
\definecolor{shadecolor}{rgb}{1, 1, 1}\color{fgcolor}\begin{kframe}
\begin{verbatim}
R> measles_globals <- spatPomp_Csnippet("
+    const double V[6][6] = {
+    {0,2.42,0.950,0.919,0.659,0.786},
+    {2.42,0,0.731,0.722,0.412,0.590},
+    {0.950,0.731,0,1.229,0.415,0.432},
+    {0.919,0.722,1.229,0,0.638,0.708},
+    {0.659,0.412,0.415,0.638,0,0.593},
+    {0.786,0.590,0.432,0.708,0.593,0}
+    }; 
+  ")
\end{verbatim}
\end{kframe}
\end{knitrout}
We now construct a Csnippet for initializing the latent process at time $t_0$.
This is done using unit-specific IVPs, as discussed in Sections~\ref{subsec:init} and~\ref{sec:params}.
Here, the IVPs are \code{S1\_0}$,\dots,$\code{S6\_0}, \code{E1\_0},$\dots$,\code{E6\_0}, and \code{I1\_0},$\dots$,\code{I6\_0}.
These code for the initial value of the corresponding states, \code{S1},$\dots$,\code{S6}, \code{E1},$\dots$,\code{E6}, and \code{I1},$\dots$,\code{I6}.
Additional book-keeping states, \code{C1},$\dots$,\code{C6}, count accumulated cases during an observation interval and so are initialized to zero.
The arguments \code{unit\_ivpnames = c('S','E','I')} and \code{unit\_statenames = c('S','E','I','C')} enable \code{spatPomp()} to expect these variables and define then as needed when compiling the Csnippets.
Similarly, \code{unit\_covarnames = 'P'} declares the corresponding unit-specific population covariate.
This is demonstrated in the following Csnippet specifying \code{rinit}.
\begin{knitrout}
\definecolor{shadecolor}{rgb}{1, 1, 1}\color{fgcolor}\begin{kframe}
\begin{verbatim}
R> measles_rinit <- spatPomp_Csnippet(
+    unit_statenames = c('S','E','I','C'),
+    unit_ivpnames = c('S','E','I'),
+    unit_covarnames = c('P'),
+    code = "
+      for (int u=0; u<U; u++) {
+        S[u] = round(P[u]*S_0[u]);
+        E[u] = round(P[u]*E_0[u]);
+        I[u] = round(P[u]*I_0[u]);
+        C[u] = 0;
+      }
+    "
+  )
\end{verbatim}
\end{kframe}
\end{knitrout}
The \code{rprocess} Csnippet has to encode only a rule for a single Euler increment from the process model.
\proglang{C} definitions are provided by \pkg{\spatPomp} for all parameters, state variables, covariates, \code{t}, \code{dt} and \code{U}.
Any additional variables required must be declared as \proglang{C} variables within the Csnippet.
\begin{knitrout}
\definecolor{shadecolor}{rgb}{1, 1, 1}\color{fgcolor}\begin{kframe}
\begin{verbatim}
R> measles_rprocess <- spatPomp_Csnippet(
+    unit_statenames = c('S','E','I','C'),
+    unit_covarnames = c('P','lag_birthrate'),
+    code = "
+      double beta, seas, Ifrac, mu[7], dN[7];
+      int u, v;
+      int BS=0, SE=1, SD=2, EI=3, ED=4, IR=5, ID=6;
+  
+      beta = R0*(muIR+muD);
+      t = (t-floor(t))*365.25;
+      seas = (t>=7&&t<=100)||(t>=115&&t<=199)||(t>=252&&t<=300)||(t>=308&&t<=356)
+        ? 1.0 + A * 0.2411/0.7589 : 1.0 - A;
+  
+      for (u = 0 ; u < U ; u++) {
+        Ifrac = I[u]/P[u];
+        for (v=0; v < U ; v++) if(v != u)
+          Ifrac += g * V[u][v]/P[u] * (I[v]/P[v] - I[u]/P[u]);
+  
+        mu[BS] = lag_birthrate[u];   
+        mu[SE] = beta*seas*Ifrac*rgammawn(sigmaSE,dt)/dt; 
+        mu[SD] = muD;                
+        mu[EI] = muEI;             
+        mu[ED] = muD;  
+        mu[IR] = muIR; 
+        mu[ID] = muD;  
+  
+        dN[BS] = rpois(mu[BS]*dt);
+        reulermultinom(2,S[u],&mu[SE],dt,&dN[SE]);
+        reulermultinom(2,E[u],&mu[EI],dt,&dN[EI]);
+        reulermultinom(2,I[u],&mu[IR],dt,&dN[IR]);
+  
+        S[u] += dN[BS] - dN[SE] - dN[SD];
+        E[u] += dN[SE] - dN[EI] - dN[ED];
+        I[u] += dN[EI] - dN[IR] - dN[ID];
+        C[u] += dN[EI];           
+      }
+    "
+  )
\end{verbatim}
\end{kframe}
\end{knitrout}
The measurement model is chosen to allow for overdispersion relative to the binomial distribution with success probability $\rho$.
Here, we show the Csnippet defining the unit measurement model.
The \code{lik} variable is pre-defined and is set to the evaluation of the unit measurement density in either the log or natural scale depending on the value of \code{give\_log}.
\begin{knitrout}
\definecolor{shadecolor}{rgb}{1, 1, 1}\color{fgcolor}\begin{kframe}
\begin{verbatim}
R> measles_dunit_measure <- spatPomp_Csnippet("
+    double m = rho*C;
+    double v = m*(1.0-rho+psi*psi*m);
+    lik = dnorm(cases,m,sqrt(v),give_log);
+  ")
\end{verbatim}
\end{kframe}
\end{knitrout}
The user may also directly supply \code{dmeasure} that returns the product of unit-specific measurement densities.
The latter is needed to apply \pkg{pomp} functions which require \code{dmeasure} rather than \code{dunit\_measure}.
We create the corresponding Csnippet in \code{measles\_dmeasure}, but do not display the code here.
Next, we construct a Csnippet to code \code{runit\_measure},

\begin{knitrout}
\definecolor{shadecolor}{rgb}{1, 1, 1}\color{fgcolor}\begin{kframe}
\begin{verbatim}
R> measles_runit_measure <- spatPomp_Csnippet("
+    double cases;
+    double m = rho*C;
+    double v = m*(1.0-rho+psi*psi*m);
+    cases = rnorm(m,sqrt(v));
+    if (cases > 0.0) cases = nearbyint(cases);
+    else cases = 0.0;
+  ")
\end{verbatim}
\end{kframe}
\end{knitrout}

We also construct, but do not display, a Csnippet \code{measles\_rmeasure} coding the \class{pomp} version \code{rmeasure}.
Next, we build Csnippets for \code{eunit\_measure} and \code{vunit\_measure} which are required by EnKF and IEnKF.
These have defined variables named \code{ey} and \code{vc} respectively, which should return  $\E[Y_{\unit,\time}\given X_{\unit,\time}]$ and $\var[Y_{\unit,\time}\given X_{\unit,\time}]$.
For our measles model, we have 
\begin{knitrout}
\definecolor{shadecolor}{rgb}{1, 1, 1}\color{fgcolor}\begin{kframe}
\begin{verbatim}
R> measles_eunit_measure <- spatPomp_Csnippet("ey = rho*C;")
R> measles_vunit_measure <- spatPomp_Csnippet("
+    double m = rho*C;
+    vc = m*(1.0-rho+psi*psi*m);
+  ")
\end{verbatim}
\end{kframe}
\end{knitrout}

It is convenient (but not necessary) to supply a parameter vector of scientific interest for testing the model.
Here, we use a parameter vector with duration of infection and latent period both set equal to one week, following \citet{xia04}, and the basic reproduction number set to $\Rzero{ } =30$.
The gravitational constant, $g=1500$, was picked by qualitative visual matching of simulations.
\begin{knitrout}
\definecolor{shadecolor}{rgb}{1, 1, 1}\color{fgcolor}\begin{kframe}
\begin{verbatim}
R> IVPs <- rep(c(0.032,0.00005,0.00004,0.96791),each=6) 
R> names(IVPs) <- paste0(rep(c('S','E','I','R'),each=6),1:6,"_0")
R> measles_params <- c(R0=30,A=0.5,muEI=52,muIR=52,muD=0.02,
+    alpha=1,sigmaSE=0.01,rho=0.5,psi=0.1,g=1500,IVPs)
\end{verbatim}
\end{kframe}
\end{knitrout}
Special treatment is afforded to latent states that track accumulations of other latent states between observation times.
These accumulator variables should be reset to zero at each observation time.
The \code{unit\_accumvars} argument provides a facility to specify the unit-level names of accumulator variables, extending the \code{accumvars} argument to \code{pomp()}.
Here, there is one accumulator variable, \code{C}, which is needed since each case report corresponds to new reported infections accumulated over a measurement interval.
The pieces of the SpatPOMP are now added to \code{measles6} via a call to \code{\spatPomp}:
\begin{knitrout}
\definecolor{shadecolor}{rgb}{1, 1, 1}\color{fgcolor}\begin{kframe}
\begin{verbatim}
R> measles6 <- spatPomp(
+    data = measles6,
+    covar = measles_covar,
+    unit_statenames = c('S','E','I','R','C'),
+    unit_accumvars = c('C'),
+    paramnames = names(measles_params),
+    rinit = measles_rinit,
+    rprocess = euler(measles_rprocess, delta.t=1/365),
+    dunit_measure = measles_dunit_measure,
+    eunit_measure = measles_eunit_measure,
+    vunit_measure = measles_vunit_measure,
+    runit_measure = measles_runit_measure,
+    dmeasure = measles_dmeasure,
+    rmeasure = measles_rmeasure,
+    globals = measles_globals
+  )
\end{verbatim}
\end{kframe}
\end{knitrout}
Here, we have not filled the \code{skeleton} and \code{munit\_measure} arguments, used by \code{girf} and \code{abfir}.
These can be found in the \pkg{spatPomp} package source code for \code{measles()}.

In Figure~\ref{fig:measles_sim_plot}, we compare a simulation from \code{measles6} with the data.
Epidemiological settings may be clearer when looking on the log scale, and so we use the \code{log=TRUE} argument to \code{plot()}.
This figure shows some qualitative similarity between the simulations and the data, with opportunity for future work to investigate discrepancies.

\begin{knitrout}
\definecolor{shadecolor}{rgb}{1, 1, 1}\color{fgcolor}\begin{figure}

{\centering \includegraphics[width=0.9\linewidth]{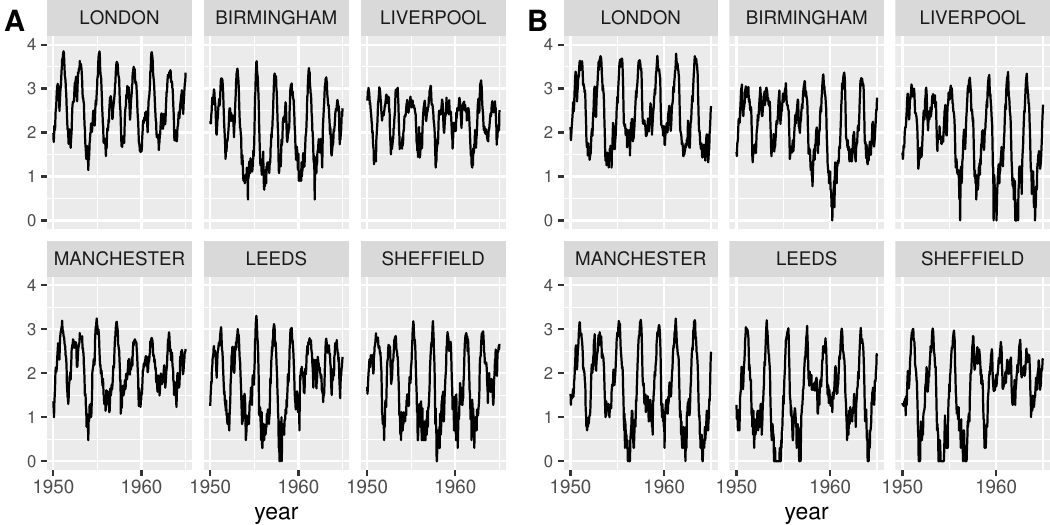} 

}

\caption{A: reported measles cases in two week intervals for the six largest cities in England, \code{plot(measles6,log=TRUE)}. B: simulated data, \code{plot(simulate(measles6),log=TRUE)}. The vertical scale is \code{log10(cases+1)}.}\label{fig:measles_sim_plot}
\end{figure}

\end{knitrout}

\section{Conclusion}
\label{sec:conclusion}

The \pkg{\spatPomp} package is both a tool for data analysis based on SpatPOMP models and a principled computational framework for the ongoing development of inference algorithms.
Although \pkg{\spatPomp} development has focused on algorithms with the plug-and-play property, it supports the development of new algorithms with and without this property.
Current examples have emphasized biological metapopulation dynamics, but diverse applications fit into the SpatPOMP model class.
Spatiotemporal data analysis using mechanistic models is a nascent topic, and future methodological developments are anticipated.

Complex models and large datasets can challenge available computational resources.
With this in mind, key components of the \pkg{\spatPomp} package and associated models are written in \proglang{C}.
This permits competitive performance on benchmarks \citep{fitzjohn20} within an R environment.
The use of multi-core computing is helpful for computationally intensive methods.
Two common computationally intensive tasks in \pkg{spatPomp} are the assessment of Monte~Carlo variability and the investigation of the roles of starting values and other algorithmic settings on optimization routines.
These tasks require only embarrassingly parallel computations and need no special discussion here.

Practical modeling and inference for metapopulation systems, capable of handling scientifically motivated nonlinear, non-stationary stochastic models, is the last open problem of the challenges raised by \citet{bjornstad01}.
Recent studies have reiterated the scientific need for such methods \citep{becker16,li20}.
Beyond the introduction provided by this tutorial, the case studies of \citet{wheeler24} and \citet{li24} provide source code describing \pkg{spatPomp} data analysis meeting this need.

\section*{Acknowledgments}
This work was supported by National Science Foundation grants DMS-1761603 and DMS-1646108, and National Institutes of Health grants 1-U54-GM111274 and 1-U01-GM110712.
We recognize those who have participated in the development and testing of \pkg{spatPomp}, especially Allister Ho, Zhuoxun Jiang, Jifan Li, Patricia Ning, Eduardo Ochoa, Rahul Subramanian and Jesse Wheeler.

%\bibliography{bib-spatpomp}

\end{document}